%% file: main.tex
\input{venue-naturemat}
\input{preamble}
\title{Coexisting Kondo and Weyl-like transport in the kagome heavy-fermion metal \ce{Yb_yCo3Ge_{3-x}Sn_x}}
\input{authors}
\date{\today}

\begin{document}
\maketitle	

\section{Abstract}

\textbf{
Electron interactions and electronic band structures offer complementary routes to controlling electrical conduction. 
In heavy-fermion metals, Kondo hybridization produces quasiparticles with strongly enhanced masses. 
In Weyl semimetals topological band crossings create light quasiparticles and generate characteristic transport responses which can be strongly modified by Kondo renormalization. 
An open question is whether distinct transport contributions remain accessible as heavy-fermion coherence develops. 
Here we show that the kagome metal \ce{Yb_yCo3Ge_{3-x}Sn_x} retains distinguishable Kondo and Weyl-like magnetoresistance contributions below the onset of Kondo coherence. 
Magnetotransport, heat capacity and magnetization establish the correlated response, with extrapolated Sommerfeld coefficients approaching 2~J\,mol$_{\mathrm{Yb}}^{-1}$\,K$^{-2}$, while X-ray photoemission supports predominantly trivalent Yb.
Transverse-field magnetoresistance follows single-ion Kondo scaling, while longitudinal-field measurements resolve an additional negative contribution down to 0.44~K and unsaturated up to 32~T, consistent with a chiral anomaly. 
Angle-resolved photoemission reveals flat and dispersive bands, while calculations including electronic correlations and spin-orbit coupling identify kagome-derived and Yb $4f$ states near the Fermi level. 
Together, these findings support predominantly distinct electronic contributions despite hybridization. 
Their simultaneous accessibility without external Fermi-level tuning suggests a route towards combining correlated and topological conduction functionality within a single material.
}

\section{Introduction}

Electrical conduction in multiband materials reflects the combined contributions of electronic states with different orbital characters, dispersions, and scattering mechanisms. Strong correlations can reorganize these states through hybridization, changing both their individual responses and their relative contributions to transport. Resolving correlated and topological transport contributions within the shared electronic environment reveals whether their characteristic responses remain individually accessible as correlations develop. Preserving such accessibility could provide a basis for selectively addressing different conduction functions within one material, without interfaces between materials or external Fermi-level tuning.\autocite{Bernevig_etal_2022,Sun_Steglich_2013,Tanaka_Kontani_2010}

In strongly correlated heavy-fermion compounds, hybridization between localized 4$f$ states and conduction electrons produces quasiparticles with strongly enhanced effective masses that renormalize the low-energy electronic structure.\autocite{Grewe_Steglich_1991,Wirth_Steglich_2016} 
Above the Kondo temperature, incoherent scattering from local moments results in a logarithmic increase in resistivity with decreasing temperature, characteristic of the single-ion impurity Kondo effect.\autocite{Kondo_1964}
Upon further cooling, a Kondo lattice can form through the onset of coherent electron scattering across the local moments, resulting in a resistivity maximum followed by a decrease with temperature. The competition between Kondo screening and intersite magnetic interactions governs the broader low-temperature behavior.\autocite{Doniach_1977}

In topological Weyl compounds, singly-degenerate linear band crossings in the electronic structure create a solid-state analog of relativistic chiral fermions with negligible effective mass.\autocite{Armitage_etal_2018, Nielsen_Ninomiya_1983,Son_Spivak_2013} A characteristic transport signature is negative longitudinal magnetoresistance (NLMR) associated with the chiral anomaly when the magnetic field is parallel to the current (in type I Weyl systems).\autocite{Soluyanov_etal_2015} Magnetic or exchange fields can generate Weyl nodes from suitable Dirac (doubly degenerate) band crossings.\autocite{Cano_etal_2017,Hirschberger_etal_2016} 

Kondo correlations and Weyl topology can be combined in several ways. In Weyl-Kondo semimetals, Weyl states emerge from within Kondo-renormalized quasiparticle bands,\autocite{Lai_etal_2018,Grefe_etal_2020} with a giant spontaneous Hall effect reported in \ce{Ce3Bi4Pd3}.\autocite{Dzsaber_etal_2021} In \ce{YbPtBi}, conduction-band Weyl states instead produce a longitudinal chiral-anomaly contribution at elevated temperatures, but it becomes negligible on cooling as the Kondo coupling renormalizes their dispersion.\autocite{Guo_etal_2018} These examples motivate asking whether a Weyl-like contribution can remain distinguishable from the Kondo field response below the onset of coherence. 

Rare-earth kagome metals offer a natural setting in which to explore this possibility. In the \ce{AM6X6} family, a rare-earth sublattice exists within a transition-metal kagome network.\autocite{Venturini_etal_2008} The resulting band structures contain dispersive Dirac bands, flat bands and van Hove singularities characteristic of the kagome geometry\autocite{johnston_hoffmann_1990} together with weakly dispersive 4$f$ electronic bands. Studies of \ce{YbV6Sn6} and \ce{YbCr6Ge6} have established Kondo correlations and hybridization involving kagome-derived states.\autocite{Guo_etal_2023a,Lou_etal_2025b,Lee_etal_2026} \ce{Yb_yCo3Ge_{3-x}Sn_x} offers a partially occupied Yb sublattice within a Co kagome host.\autocite{Weiland_etal_2020} Earlier measurements identified a Kondo-like upturn and pressure-induced coherence,\autocite{Wang_etal_2022,Cheng_etal_2024} while calculations on its ordered structural counterpart place flat and dispersive states near the Fermi level.\autocite{Tan_etal_2024}

Here we show that \ce{Yb_yCo3Ge_{3-x}Sn_x} retains a distinguishable Weyl-like longitudinal magnetoresistance contribution down to 0.44~K, below the onset of Kondo coherence. Transport, heat capacity and magnetization establish the Yb-associated correlated response, with extrapolated Sommerfeld coefficients approaching 2~J\,mol$_{\mathrm{Yb}}^{-1}$\,K$^{-2}$. Transverse magnetoresistance and magnetization are consistent with single-ion $j = 1/2$ Kondo scaling,\autocite{Schlottmann_1983} whereas longitudinal magnetoresistance contains an additional negative contribution proportional to $B^2$ and remains unsaturated up to 32~T, consistent with a Weyl-like chiral anomaly. Unlike the chiral-anomaly signal in \ce{YbPtBi} \autocite{Guo_etal_2018}, this additional response remains accessible far below the coherence onset, indicating the Kondo lattice is not inhibiting the Weyl transport signature. Photoemission and density-functional theory (DFT) calculations including Coulomb interaction $U$ and spin--orbit coupling (SOC) identify flat and dispersive states and their orbital contributions. The combined evidence points to Kondo and Weyl-like conduction arising from distinct electronic states in a single material without external tuning. 

\section{Results}
\subsection{A compositionally tunable kagome host}

\begin{figure}
    \centering
    \includegraphics[width=\figwidth]{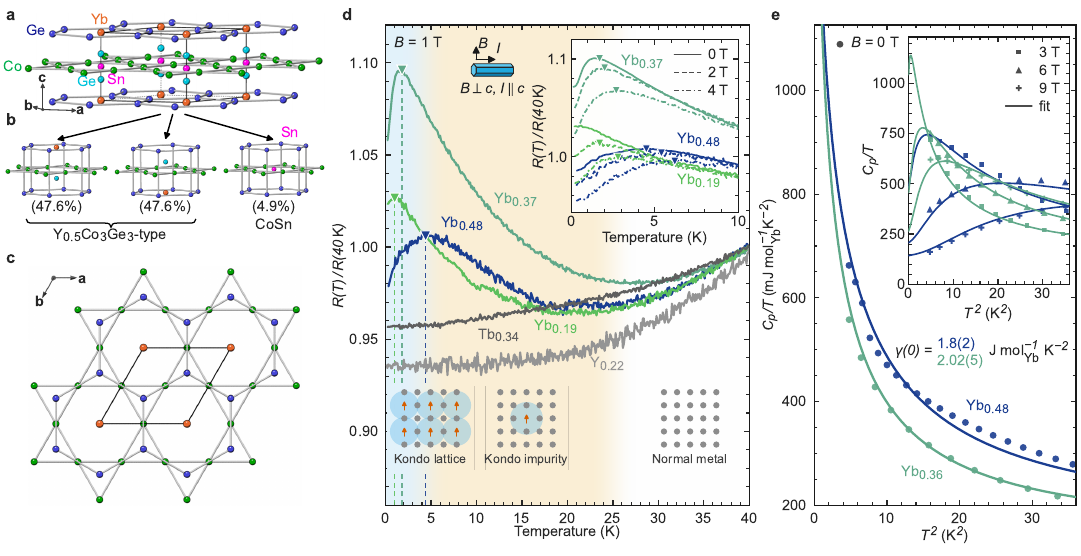}
    \caption{Crystal structure and heavy-fermion signatures of \ce{Yb_yCo3Ge_{3-x}Sn_x}.
    (a) Averaged structure of \crystal{RM154}{formula} from the single-crystal refinement ($P6/mmm$, $a = \crystal{RM154}{a}$~\AA, $c = \crystal{RM154}{c}$~\AA), with the split Ge site in cyan.
    (b) Its components, with the share of each from the refinement: the two mirror placements of the \ce{HfFe6Ge6}-type pattern, each with a Yb atom and a Ge pair in the channel, which together form the \ce{Y_{0.5}Co3Ge3} type, and the \ce{CoSn} type.
    (c) View down the $c$-axis, with the unit cell outlined.
    (d) Resistance normalized to its value at 40~K of three Yb fillings and Tb and Y analogs, at 1~T with $B \perp c$ and $I \parallel c$.
    Triangles and dashed lines mark the coherence maxima, and the shading marks the Kondo-lattice regime below the 4.4~K maximum, the Kondo-impurity regime from 4.4 to 25~K and the normal-metal regime above 25~K.
    Inset: in field the maximum moves up in temperature.
    (e) $C_p/T$ against $T^2$ at 0~T, with 3, 6 and 9~T in the inset.
    Markers are data, with errors smaller than the symbols, and solid lines are one Schotte-Schotte resonant-level fit per crystal to all ten fields from 0 to 9~T at once.}
    \label{fig:heavy-fermion}
\end{figure}

\ce{Yb_yCo3Ge_{3-x}Sn_x} combines a Co kagome network with a partially occupied Yb sublattice. Single crystals grow from a Sn flux as thin hexagonal needles along $c$ (Methods). Increasing the Yb loading of the melt raises the refined Yb occupancy,\autocite{Meduri_etal_2025} spanning $y = \crystal{RM238}{y}$ to $\crystal{RM154}{y}$ in our crystals (Supplementary Tables~\ref{stab:inventory} and \ref{stab:growth}). They adopt the disordered \ce{Yb_{0.5}Co3Ge3} structure type,\autocite{Weiland_etal_2020} with average space group $P6/mmm$ (Fig.~\ref{fig:heavy-fermion}a). Locally, the structure combines \ce{Y_{0.5}Co3Ge3}-type and \ce{CoSn}-type cells (Fig.~\ref{fig:heavy-fermion}b). Channels through the Co kagome and Ge honeycomb networks contain random sequences of occupied and empty Yb sites (Fig.~\ref{fig:heavy-fermion}c and Supplementary Fig.~\ref{sfig:structure-derivation}), so partial occupancy and occupational disorder are intrinsic features of the Yb sublattice in this structure type. This structural flexibility allows the correlated response to be examined across different Yb fillings within the same kagome structure family.

\subsection{Yb-driven heavy-fermion coherence}

Yb incorporation produces a progression from Kondo scattering to lattice coherence. All three Yb-containing crystals show a resistance upturn below approximately 25~K, consistent with the logarithmic Kondo behavior reported previously (Fig.~\ref{fig:heavy-fermion}d).\autocite{Weiland_etal_2020,Wang_etal_2022,Cheng_etal_2024} On further cooling, the resistance passes through a maximum and falls steeply, revealing coherence at ambient pressure in a compound family where this behavior was previously observed under pressure.\autocite{Cheng_etal_2024} Neither the Y nor the Tb analogue shows the corresponding upturn and coherence maximum, associating both features with Yb rather than with the Co kagome host alone.

At 1~T, applied perpendicular to $c$ to suppress the contribution from residual Sn flux, the coherence maximum decreases from 4.4~K to 1.0~K as $y$ decreases from \crystal{RM154}{y} to \crystal{RM238}{y}. The downward shift is consistent with the sensitivity of Kondo coherence to magnetic-site dilution established in other heavy-fermion systems.\autocite{Onuki_etal_1985c,Sumiyama_etal_1986,Petrovic_etal_2023,Slebarski_etal_2013,Winkler_etal_2010,Kohler_etal_2008} The maximum moves to higher temperature with field (Fig.~\ref{fig:heavy-fermion}d, inset; Supplementary Fig.~\ref{sfig:resistance-in-field}). Yb filling and Sn incorporation both vary between these crystals, making the measured trend describe the composition dependence of the coherence onset within this series.

Heat capacity independently establishes the large low-energy electronic response (Fig.~\ref{fig:heavy-fermion}e). At zero field $C_p/T$ reaches approximately $600$~mJ\,mol$_{\mathrm{Yb}}^{-1}$\,K$^{-2}$ at 2~K in both crystals, roughly 20 times above the 27~mJ\,mol$_{\mathrm{Y}}^{-1}$\,K$^{-2}$ reported for the Y analog. \autocite{Bravo_etal_2023} In perpendicular fields, the rise develops into a broad maximum that moves to higher temperature, consistent with a Schottky anomaly from a field-split doublet. We fit all ten fields from 0 to 9~T simultaneously with the Schotte--Schotte resonant-level model,\autocite{Schotte_Schotte_1975} constraining a common low-energy description through its temperature and field dependence (Supplementary Section~\ref{ssec:resonant-level}).

The resulting extrapolated Sommerfeld coefficients at $T \to 0$ are $\gamma(0) = 1.8(2)$ and $2.02(5)$ J\,mol$_{\mathrm{Yb}}^{-1}$\,K$^{-2}$ for \crystal{RM139}{short} and \crystal{RM048}{short}, respectively. These values exceed the 0.4~J\,mol$^{-1}$\,K$^{-2}$ commonly taken as the heavy-fermion threshold.\autocite{Stewart_1984a}  Across three phonon-background estimates, the entropy at 20~K remains between 0.88 and $1.07R\ln 2$ per Yb (Supplementary section~\ref{ssec:entropy}, Fig.~\ref{sfig:entropy}). This supports a ground-doublet description and argues against a substantial heat-capacity contribution from additional thermally populated crystal-field levels in this range. The field-induced maximum is instead well described by Zeeman splitting of the Kondo-broadened doublet. X-ray photoemission and Curie--Weiss magnetization independently support predominantly trivalent Yb (Supplementary section~\ref{ssec:xps} and Figs.~\ref{sfig:xps} and \ref{sfig:curie-weiss}). Together with the transport evidence, this favors a Kondo origin of the low-energy enhancement over an anomaly from excited crystal-field levels. 

\subsection{Single-ion Kondo modelling of magnetoresistance and magnetization}

\begin{figure}[b]
    \centering
    \includegraphics[width=\figwidth]{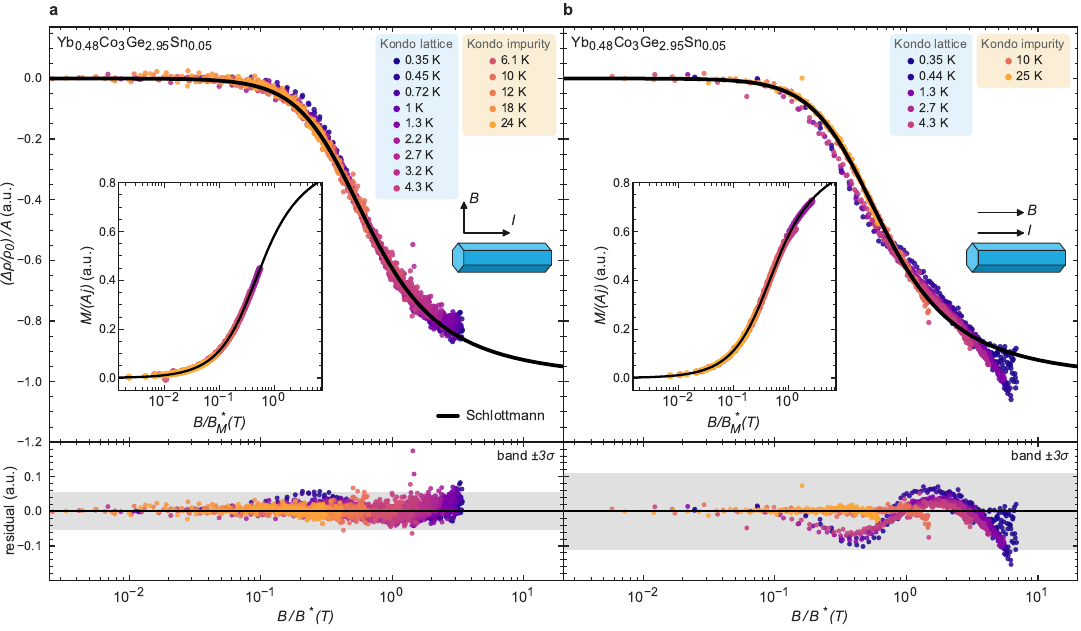}
    \caption{Single-ion Kondo scaling of the magnetoresistance and magnetization of \crystal{RM154}{formula}.
    (a) $\Delta\rho/\rho_0$ for $B \perp c$, divided by each isotherm's amplitude $A$, against $B/B^*(T)$.
    Temperatures are grouped by the regimes of Fig.~\ref{fig:heavy-fermion}d.
    The line is the $j = 1/2$ Schlottmann curve at $T = 0$. \autocite{Schlottmann_1983}
    Gray bands in the residual panels span $\pm 3\sigma$, with $\sigma$ the root-mean-square residual.
    Inset: magnetization of a crystal from the same batch, 2 to 30~K, scaled with its own $B^*_M(T)$.
    (b) The same for $B \parallel c$.
    Magnetization isotherms and their residuals are in Supplementary Fig.~\ref{sfig:mh-isotherms}.}
    \label{fig:scaling}
\end{figure}

Magnetoresistance in a transverse field follows single-ion Kondo scaling over three decades of reduced field, whereas magnetoresistance in a longitudinal field shows systematic deviations and a field scale inconsistent with magnetization (Fig.~\ref{fig:scaling}). In both configurations we measure the resistivity along $c$, using the same voltage contacts. Transverse-field MR denotes $B \perp I$ with $I \parallel c$, and longitudinal-field MR denotes $B \parallel I \parallel c$.

To test the single-ion description, we modeled the measured magnetoresistance isotherms $\Delta\rho/\rho_0$ $= \left(\rho(B) - \rho_0\right)/\rho_0$ of \crystal{RM154}{formula} between 0.35 and 25~K, with $\rho_0$ the zero-field resistivity. We used the Schlottmann model for $j = 1/2$,\autocite{Schlottmann_1983} which gives the resistivity as a universal function of $B/B^*(T)$, with a single field scale $B^*$ set by the Kondo temperature. All isotherms of one orientation are fitted jointly, with a separate amplitude for each temperature and a shared temperature-dependent field scale (Supplementary Section~\ref{ssec:coqblin-schrieffer}). This tests whether a common Kondo field dependence can account for the measurements across the impurity and coherence regimes.

For $B \perp c$ the isotherms collapse onto the Schlottmann curve over three decades of reduced field (Fig.~\ref{fig:scaling}a), with a root-mean-square residual of 2~\% of the amplitude. The characteristic scales are $B_0 = 10.1$~T and $T^* = 2.16(7)$~K. Magnetization measured on a crystal from the same batch also collapses for $B\perp c$ (inset of Fig.~\ref{fig:scaling}a), with $B_0=7(4)$~T and $T^*=2.4(26)$~K, consistent with the transport scales within the larger uncertainties imposed by the 5~T measurement range (Fig.~\ref{fig:scaling}a, inset; Supplementary Fig.~\ref{sfig:mh-isotherms}). The successful collapse extends below the 4.4~K coherence maximum: onset of lattice coherence does not immediately eliminate single-ion scaling of the field response. Small low-field deviations emerging below 0.72~K are not captured by this model (Supplementary Section~\ref{ssec:coqblin-schrieffer}).

For $B \parallel c$, the nominal Kondo temperature remains similar, $T^* \approx 2.4$~K, while $B_0$ falls to approximately 5.0~T. An anisotropic doublet could account for different field scales, but the fit residual increases to 4~\% and develops systematic structure (Fig.~\ref{fig:scaling}b and Supplementary Section~\ref{ssec:coqblin-schrieffer}). More decisively, the magnetization does not agree; along $c$ it has a Curie-like scale $B_M^* \propto T$: at 2~K, $B_M^* = 1.9$~T, compared with $B^* = 6.5$~T from MR, a factor of 3 disagreement. 
Thus, the magnetic moments respond on a field scale more than three times smaller, additionally implying the longitudinal-field MR cannot be fully attributed to Kondo scattering.

Changing the single-ion anisotropy therefore does not reconcile the magnetic and resistive responses for $B \parallel c$. The mismatch motivates testing for an additional negative MR contribution, particularly at fields beyond the response associated with polarization of the doublet. Supplementary Sections~\ref{ssec:coqblin-schrieffer} and \ref{ssec:g-factors} give the fit uncertainties and comparisons among the independently inferred field scales.

\subsection{Non-saturating negative longitudinal magnetoresistance}

\begin{figure}[b]
    \centering
    \includegraphics[width=\figwidth]{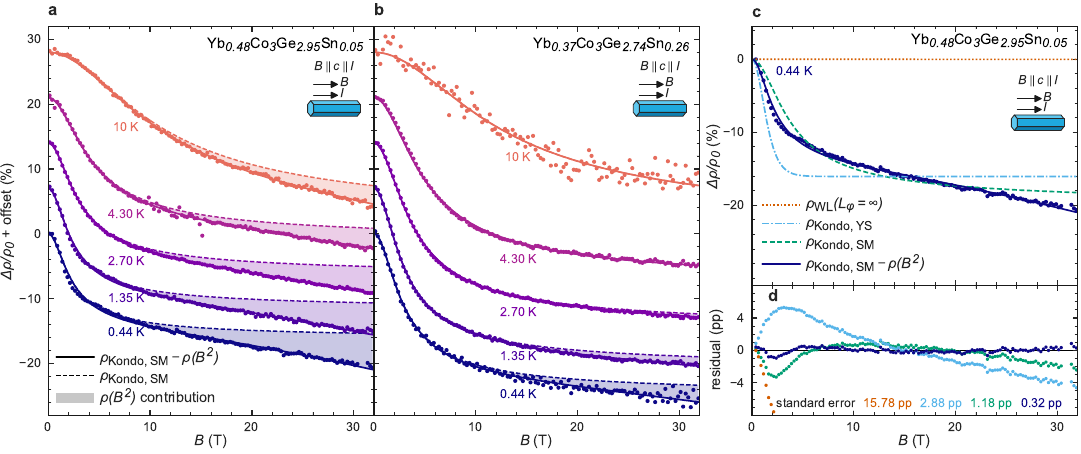}
    \caption{Kondo and extra-channel parts of the high-field magnetoresistance for $B \parallel c \parallel I$.
    (a) \crystal{RM154}{formula} between 0.44 and 10~K, offset by 7 percentage points per isotherm.
    Markers are data.
    Solid lines are fits of each isotherm from 0.3 to 32~T with the $j = 1/2$ Schlottmann resistivity $\rho_\mathrm{Kondo,SM}$ \autocite{Schlottmann_1983} and an extra channel in parallel, dashed lines are $\rho_\mathrm{Kondo,SM}$ of the same fit, and the shading between them is $\rho(B^2)$, the resistivity the additional channel removes.
    The 25~K isotherm is left out, because its Kondo field scale is larger than 32~T and the fit cannot separate the two parts at that temperature.
    The schematic shows the field and the current along the needle axis $c$.
    (b) The same for \crystal{RM230}{formula}.
    At 4.3 and 10~K the fit resolves no channel, so the solid line is the Schlottmann curve alone.
    (c) The 0.44~K isotherm of (a) against $\rho_\mathrm{Kondo,SM}$ fitted alone without the channel, against three-dimensional weak localization $\rho_\mathrm{WL}$ \autocite{Kawabata_1980} at an infinite phase-coherence length, and against Yosida's spin-disorder term $\rho_\mathrm{Kondo,YS}$ \autocite{Yosida_1957} with its field scale fixed by the low-field data, the last two without $\rho_\mathrm{Kondo,SM}$.
    The solid line is the fit of (a), $\rho_\mathrm{Kondo,SM} - \rho(B^2)$.
    (d) Residuals of the fits in (c) in percentage points (pp), with the standard error of each fit.
    The weak-localization residual leaves the panel above about 2~T.
    }
    \label{fig:high-field}
\end{figure}
High magnetic fields expose a longitudinal-field MR contribution beyond the single-ion Kondo description. At 0.44~K, the MR of \crystal{RM154}{formula} reaches $-21\,\%$ at 32~T and continues to decrease (Fig.~\ref{fig:high-field}a,c), whereas the fitted Schlottmann response levels off above approximately 20~T (Supplementary Section~\ref{ssec:high-field}, Figs.~\ref{fig:high-field}c,d). Fig.~\ref{fig:high-field}b shows the same measurement on \crystal{RM230}{formula}, which has a lower Yb filling. The distinction is visible in the field dependence: the measured resistivity keeps falling beyond the response captured by the Kondo model.

We describe this behavior by fitting each longitudinal-field isotherm with a Schlottmann contribution whose amplitude $A$ and field scale $B^*$ vary freely, together with an additional parallel channel with quadratic field dependence (Supplementary Section~\ref{ssec:extra-channel}). Figure~\ref{fig:high-field} denotes the associated resistivity reduction by $\rho(B^2) = \rho_\mathrm{Kondo,SM} - \rho$. Including this channel reduces the residual standard error of the 0.44~K fit for \crystal{RM154}{formula} from 1.18 to 0.32 percentage points (Fig.~\ref{fig:high-field}d), a 73\% reduction and a factor-of-3.7 improvement. These values quantify the fit residuals, rather than the fractional change in sample resistance. The improved agreement removes the systematic high-field mismatch while retaining a clearly discernible Kondo contribution with independently refined parameters.

Crucially, the additional contribution remains resolved down to 0.44~K in both crystals, below their low-field coherence temperatures. In \crystal{RM154}{formula}, it is resolved from 0.44 to 10~K and weakens on warming. In \crystal{RM230}{formula}, it is weaker and is resolved only up to 2.7~K; at 4.3 and 10~K the fits do not separate an additional channel. Negative MR persists to higher temperature, but at 25~K the Kondo field scale exceeds the accessible field range. The result is therefore their simultaneous visibility at low temperature, rather than the sign of the total MR alone.

The composition dependence provides a further comparison. The characteristic channel field $B_\mathrm{ch}$, for which a smaller value denotes a stronger contribution, rises from 110 to 148~T between 0.44 and 10~K in \crystal{RM154}{formula}. In \crystal{RM230}{formula}, it rises from 149 to 338~T between 0.44 and 2.7~K. The same functional description thus captures different contribution strengths and temperature dependences in the two compositions. \todo{look into Cw vs Bch}

Alternative negative-MR mechanisms fail to reproduce the observed magnitude and field dependence within the models tested. Three-dimensional weak localization\autocite{Kawabata_1980} gives approximately $-0.02\,\%$ at 32~T, about a thousand times smaller than the measured response. The spin-disorder term (related to e.g. paramagnetic impurities) proportional to $-M^2$ and constrained by low-field MR and the measured Curie $g$ factor \autocite{Yosida_1957}, saturates and misses the high-field decrease (Fig.~\ref{fig:high-field}c). Allowing these models to vary freely requires unphysical parameters or leaves systematic residuals in both crystals (Supplementary Figs.~\ref{sfig:model-comparison} and \ref{sfig:model-comparison-rm230}). 
Furthermore, contact-pair measurements disfavor the current-jetting scenario considered in Supplementary Section~\ref{ssec:current-jetting}.
Among the descriptions tested, only the Kondo response plus the additional negative MR channel accounts for the full field range. Its field dependence is consistent with the chiral-anomaly contribution expected in Weyl systems,\autocite{Son_Spivak_2013} motivating comparison with the electronic structure.

\subsection{Dispersive kagome bands and Yb-derived flat states}

\begin{figure*}
    \centering
    \resizebox{\textwidth}{!}{%
    \includegraphics[height=5cm]{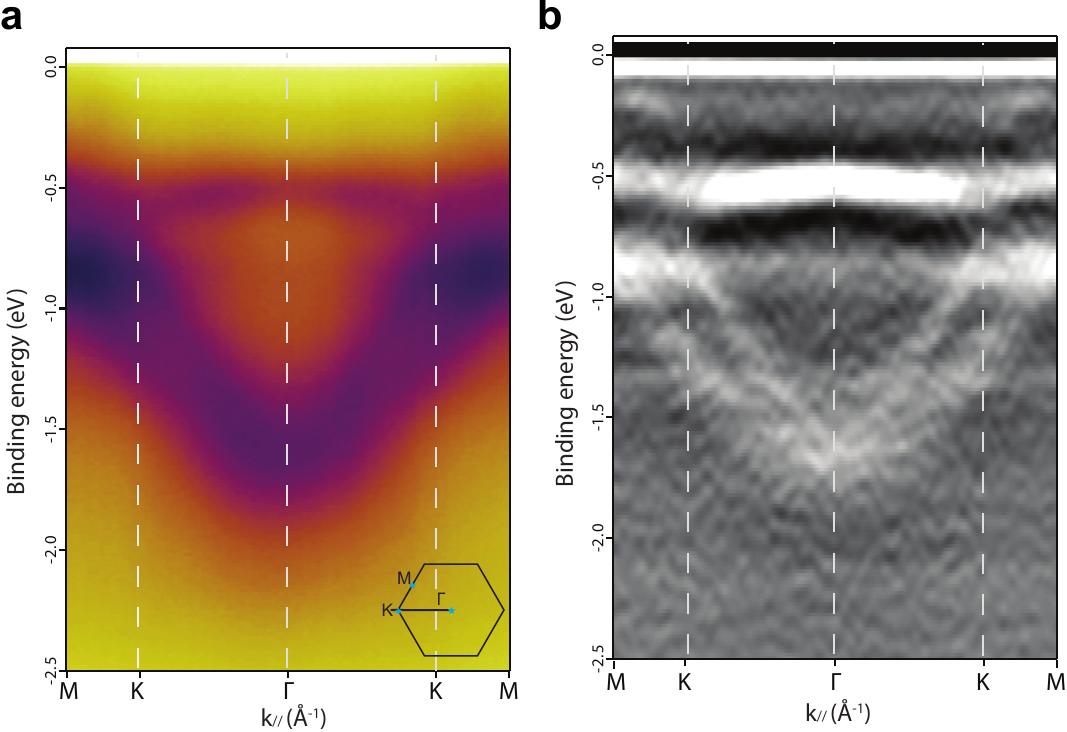}%
    \hspace{2mm}%
    \includegraphics[height=5cm]{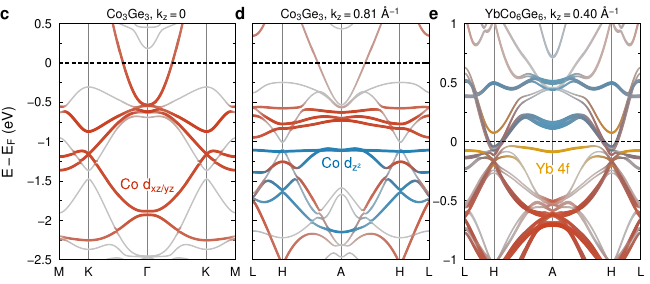}%
    }\par\vspace{2mm}
    \includegraphics[width=\textwidth]{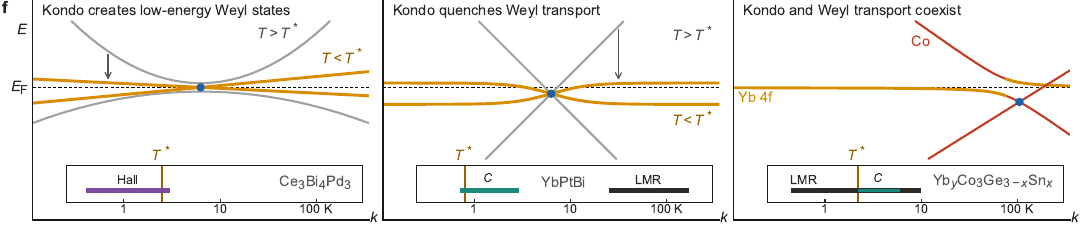}
    \caption{\ce{Yb_yCo3Ge_{3-x}Sn_x} band structure and Weyl--Kondo scenarios.
    (a) Photoemission intensity map of \crystal{RM154}{formula} taken along the $\Gamma$-K-M high-symmetry direction.
    (b) Second derivative of the data in (a), highlighting the flat and linear bands.
    (c-e) Density-functional theory (DFT) calculations of Co3Ge3 (c,d) and \ce{Yb_{0.5}Co3Ge3} (e). Red, blue, and orange indicate Co $d_{xz/yz}$, Co $d_{z^2}$ and Yb $4f$ orbital contributions, respectively.
    (f) Three schematic relationships between Kondo correlations and Weyl states.
    Grey bands are above $T^*$, orange bands are the heavy $4f$-derived states below it, and blue and red dots are Weyl nodes of opposite chirality.
    Left: Kondo hybridization and band renormalization creates heavy Weyl nodes at $E_\mathrm{F}$, so their signatures appear only below $T^*$.\autocite{Lai_etal_2018}
    Middle: Pre-existing Weyl nodes become strongly renormalized below $T^*$ turning heavy and quenching their transport signature.
    Right: proposed interpretation in which predominantly distinct electronic states allow Kondo and Weyl like transport contributions to remain distinguishable  through $T^*$ despite hybridization.
    The insets give an example of each scenario, showing the temperature ranges in which its signatures are observed, with $T^*$ as a vertical line: the spontaneous Hall effect of \ce{Ce3Bi4Pd3},\autocite{Dzsaber_etal_2021} the cubic specific heat $C$ and the negative longitudinal magnetoresistance (LMR) of \ce{YbPtBi},\autocite{Guo_etal_2018} and $C$ and the LMR of \ce{Yb_yCo3Ge_{3-x}Sn_x} from this work.
    }
    \label{fig:arpes}
\end{figure*}

The electronic-structure calculation results identify dispersive bands in the Co kagome host before Yb is introduced, while including Yb supplies near-Fermi-level $4f$ states and hybridization. Angle-resolved photoemission spectroscopy (ARPES) on \crystal{RM154}{formula} shows both dispersive and flat bands (Fig.~\ref{fig:arpes}a,b). Approximately linear bands extend towards the Fermi level, while several flat features occur at higher binding energies. The second derivative highlights these dispersions. These observations establish the coexistence of different band shapes; their orbital assignments and the detailed near-Fermi-level structure are examined through calculations.

We first calculate the CoGe reference host in its primitive \ce{CoSn}-type cell, \ce{Co3Ge3}, using DFT with spin--orbit coupling (SOC) (Fig.~\ref{fig:arpes}c,d). Its Co $d_{xz}/d_{yz}$ and $d_{z^2}$ orbitals form the characteristic kagome dispersions including Dirac-like bands crossing at K and a nearly flat band, at $k_z = 0$ and $k_z = \pi/c$ respectively (Fig.~\ref{fig:arpes}c,d). Here $c$ denotes the primitive CoGe lattice parameter. The Yb-containing calculation uses an ordered arrangment in a cell doubled along the $c$ axis like in \ce{Yb_{0.5}Co3Ge3}, which folds both $k_z$ onto $k_z = 0$, so with Yb randomly distributed ARPES should and does find a superposition of the two bandstructures (Supplementary Section~\ref{ssec:dft}).

At $k_z = \pi/2c$, where the kagome mirror no longer separates them, the two orbital sets hybridize into Dirac-like bands with a crossing near the Fermi level and an approximately 50~meV gap with spin-orbit coupling. In \ce{Yb_{0.5}Co3Ge3} this plane becomes the zone boundary, and a magnetic DFT+$U$+SOC solution places a flat Yb $4f$ band at the Fermi level that crosses and hybridizes with these Dirac bands around H (Fig.~\ref{fig:arpes}e). These calculations establish that dispersive kagome states coexist with near-Fermi-level Yb $4f$ states. Combined with the transport decomposition, they support a picture in which different electronic states contribute predominantly to the Kondo and Weyl-like responses.

\section{Discussion}
Our main result is the persistence of a distinguishable Weyl-like longitudinal-field MR contribution below the low-field onset of Kondo coherence. Thermodynamics and temperature-dependent transport establish the correlated response. Transverse-field MR follows single-ion Kondo scaling, whereas comparison with magnetization exposes a mismatch between the longitudinal-field scales inferred from the two measurements. At high fields, the resistivity continues to decrease beyond the fitted Kondo response. The Kondo amplitude and field scale are refined independently for each isotherm, yet a discernible Kondo contribution remains alongside the additional Weyl-like channel. Together with the independently established correlated state, this separability supports predominantly distinct electronic origins for the two responses.

Figure~\ref{fig:arpes}f contrasts three relationships between Kondo correlations and Weyl states. First, Kondo hybridization and renormalization can generate low-energy Weyl quasiparticles, as proposed for Weyl--Kondo semimetals.\autocite{Lai_etal_2018,Grefe_etal_2020} In \ce{Ce3Bi4Pd3}, a spontaneous Hall response emerges in the coherent regime, but a longitudinal chiral-anomaly contribution is not resolved.\autocite{Dzsaber_etal_2021} Second, Kondo coupling can strongly renormalize bands that already host Weyl nodes. In the interpretation proposed for \ce{YbPtBi}, the reduced Fermi velocity makes the longitudinal chiral-anomaly contribution negligible on cooling, although Hall and thermodynamic signatures remain.\autocite{Guo_etal_2018} Third, different electronic states can contribute predominantly to Kondo and Weyl-like transport, allowing both characteristic responses to remain distinguishable as coherence develops.

Our results support this third interpretation. The additional longitudinal-field contribution remains resolved down to 0.44~K, and the high-field fits distinguish it from the Kondo response. ARPES reveals flat and dispersive features, while our calculations identify host-derived dispersive states alongside Yb-derived states and their hybridization. Together with earlier calculations on the related Co kagome structure,\autocite{Tan_etal_2024} these observations provide a possible electronic basis for the distinct transport contributions that does not require vanishing hybridization. Future Hall measurements could provide complementary information on the carrier contributions and their evolution through Kondo coherence, however the needle-like crystal morphology makes these measurements challenging.

The structural flexibility of the \ce{A_{0.5}M3X3} family provides a route to exploring the conditions that preserve distinct correlated and topological transport contributions. Rare-earth occupancy and substitution can modify the local-moment subsystem, while transition-metal substitution changes the host electronic structure and magnetic behavior.\autocite{Weiland_etal_2020,Meduri_etal_2025,Fau_etal_2026} Members of this family exhibit ferromagnetic-like, antiferromagnetic or ferrimagnetic order,\autocite{Meduri_etal_2025,Li_etal_2025a,Fau_etal_2026} and \ce{Y_{0.5}Fe3Sn3} displays an anomalous Hall contribution beyond conventional magnetization scaling.\autocite{Liu_etal_2023b,Ding_etal_2026} Related structure types also accommodate channel-site elements as different as Ca and U.\autocite{Braun_Hlukhyy_2023,Goncalves_etal_1994} This chemical flexibility makes the family a promising setting in which to identify further Kondo systems with distinguishable Weyl-like transport.

Taken together, magnetotransport, heat capacity, magnetization, photoemission and electronic-structure calculations identify \ce{Yb_yCo3Ge_{3-x}Sn_x} as a kagome heavy-fermion metal with distinguishable Kondo and Weyl-like transport contributions. The additional negative longitudinal-field MR remains resolved below the low-field onset of Kondo coherence and continues to grow up to 32~T, consistent with a chiral-anomaly contribution.\autocite{Nielsen_Ninomiya_1983,Son_Spivak_2013} Its separability from the Kondo response, together with the electronic-structure results, supports predominantly distinct electronic origins despite hybridization. This interpretation extends the search for correlated topological transport beyond Weyl states formed within Kondo-renormalized bands.\autocite{Lai_etal_2018,Dzsaber_etal_2021,Grefe_etal_2020} The simultaneous accessibility of both responses without external Fermi-level tuning provides a setting for investigating how different electronic states retain their characteristic contributions to conduction. Establishing how their relative weights can be controlled could enable selective access to correlated and topological conduction functionality within a single material.

\section{Methods}

\subsection{Crystal growth and structure}
Single crystals of \ce{Yb_yCo3Ge_{3-x}Sn_x} were grown using the flux growth method.\autocite{Canfield_Fisk_1992}
The sample \crystal{RM154}{formula} was grown by combining Yb (99.9~\% ingots), Co (99.998~\% powder), Ge (99.999~\% pieces) and Sn (99.99~\% shots) into an alumina Canfield crucible \autocite{Canfield_etal_2016} in a ratio of 3:2:7:52, respectively, and sealed under 1/3~atm of Ar gas.
The crucible was then heated to 1175~$^\circ$C from 300~$^\circ$C at a rate of 100~$^\circ$C/h in a programmable furnace.
The mixture dwelled here for 24~h and then cooled to 815~$^\circ$C at a rate of 3~$^\circ$C/h.
The crucible was subsequently removed from the furnace, inverted and centrifuged at 1500~rpm to separate the single crystals from the flux.
Residual Sn flux on the surface was etched using a 1:1 ratio of \ce{H2O} and \ce{HCl} by volume.
The reaction yields thin hexagonal needles, which can be connected to the amount of disorder across the $ab$ plane promoting crystallization mainly along the $c$ axis.
Larger crystals were obtained by lowering the amount of Sn using the ratio 3:2:7:20.
To tune the Yb occupancy of \ce{Yb_yCo3Ge_{3-x}Sn_x}, a Yb-deficient stoichiometric ratio was employed.
The \crystal{RM048}{short} and \crystal{RM230}{short} samples were grown by combining the elements in a ratio of 1.2:6:6:25 and 1:6:6:25, respectively.
The \crystal{RM238}{short} sample was grown by lowering the amount of Sn flux using the ratio 1:6:6:10.
We hypothesized that rather than the Sn acting purely as the flux, lowering the relative stoichiometry allows Sn to behave more as a reactant.
This explains the increased Sn incorporation on the \ce{CoSn} site, inducing more occupancy of that structure type relative to the \ce{Y_{0.5}Co3Ge3} structure type.
The \crystal{RM176}{short} sample was grown using Y (99.9~\% pieces), Co, Ge and Sn in a ratio of 1:6:6:20, respectively, sealed and heated to 1175~$^\circ$C.
The reaction dwelled here for 6~h before being cooled to 500~$^\circ$C at a rate of 5~$^\circ$C/h. \autocite{Bravo_etal_2023}
The \crystal{NJF043}{short} sample was grown using Tb (99.9~\% ingots), Co, Ge and Sn in a ratio of 1.2:6:6:20, respectively, sealed and heated to 1100~$^\circ$C.
The reaction dwelled here for 24~h before being cooled to 980~$^\circ$C at a rate of 50~$^\circ$C/h.
The reaction underwent an additional cooling step to 940~$^\circ$C at a rate of 3~$^\circ$C/h and dwelled here for 72~h. \autocite{Fau_etal_2026}
Both reactions were inverted and centrifuged at 1500~rpm and etched with 1:1 \ce{H2O}:\ce{HCl} to remove residual Sn flux.

The transport and magnetization of each filling were measured on different crystals of one growth batch, which have the same Yb content within 0.01.
The heat capacity was measured on crystals of other batches with the same filling as the transport crystals.
The structure, the Yb occupancy, the split Ge site and the Sn content of each batch were refined from single-crystal X-ray diffraction at room temperature.

\subsection{Transport}
The resistance was measured with a four-probe lock-in technique, with a current of 100 to 200~$\mu$A along the $c$-axis.
The contact spacings were measured through a microscope.
The voltage contacts V56 on \crystal{RM154}{formula} used in the main text are 230~$\mu$m apart, and the second pair V12 is 1100~$\mu$m apart.
Below 4~K the temperature during a field sweep was held by a constant pressure over the pumped \ce{^4He} or \ce{^3He} bath, and above 4~K by a fixed heater output.
Fields up to 32~T were applied along $c$ ($B \parallel c \parallel I$) and perpendicular to it ($B \perp c$), on the same contacts.
Each isotherm is symmetrized in field and binned as described in Supplementary Section~\ref{ssec:mr-conventions}.

\subsection{Magnetization and heat capacity}
The magnetization of \crystal{RM154}{short} (0.17~mg) and \crystal{RM230}{short} (0.34~mg) crystals was measured in a Quantum Design MPMS-XL magnetometer in DC mode between 1.8 and 300~K in fields up to 5~T, after zero-field cooling, for both field orientations, with the $c$-axis aligned to the field by eye to better than a degree.
The heat capacity of \crystal{RM139}{short} (16.2~mg) and \crystal{RM048}{short} (5.9~mg) was measured in a Quantum Design PPMS from 2.2~K in fields from 0 to 9~T perpendicular to $c$, and at zero field up to 200 and 120~K.

\subsection{Angle-resolved photoemission}
ARPES measurements were performed at the Bloch beamline (MAX IV Laboratory).
Data was acquired at 18~K, with a photon energy of $h\nu = 188$~eV using linear horizontally polarized light.
Samples were mounted on a copper plate using silver epoxy and cleaved in situ prior to measurements.
ARPES data were analyzed using WaveMetrics Igor Pro 9 software.

\subsection{Density-functional calculations}
Density-functional calculations were performed with the Vienna Ab initio Simulation Package (VASP 6.4.3) \autocite{Kresse_Furthmuller_1996,Kresse_Furthmuller_1996a} using the projector-augmented-wave (PAW) method \autocite{Blochl_1994,Kresse_Joubert_1999} and the generalized-gradient approximation of Perdew, Burke and Ernzerhof (PBE) \autocite{Perdew_etal_1996}.
The PAW datasets treat Yb $5s^2 5p^6 4f\,6s\,5d$ (24 valence electrons, $4f$ in the valence), Co $3s^2 3p^6 3d\,4s$ (17 electrons) and Ge $3d^{10} 4s^2 4p^2$ (14 electrons) as valence states, with a plane-wave cutoff of 550~eV.
Spin-orbit coupling is included self-consistently in a noncollinear formalism.\autocite{Hobbs_etal_2000,Steiner_etal_2016}
The local Coulomb interaction in the Yb $4f$ shell is treated within the rotationally invariant DFT+$U$ scheme \autocite{Liechtenstein_etal_1995} with $U = 6$~eV and $J = 1$~eV; no $U$ is applied to Co $3d$.
\ce{Yb_{0.5}Co3Ge3} is modelled in the hexagonal \ce{HfFe6Ge6}-type structure (space group $P6/mmm$, 13 atoms per unit cell) with $a = 5.075$~\AA{} and $c = 7.804$~\AA{}\todo{structure reference [9]}.
Brillouin-zone integrations use a $\Gamma$-centred $8\times8\times6$ $k$-point mesh and Gaussian smearing of 0.05~eV, and total energies are converged to $10^{-8}$~eV.
Because density-mixing schemes did not converge once the Yb $4f$ states approach the Fermi level, the self-consistency cycle uses direct band-by-band minimization.
All input files were generated with the Atomic Simulation Environment.\autocite{HjorthLarsen_etal_2017}

\subsection{Fitting}
All fits are least-squares fits with a soft-$\ell_1$ loss, and the errors of jointly fitted parameters are jackknife errors over the isotherms or fields.
The fit functions are given in Supplementary Sections~\ref{ssec:resonant-level}, \ref{ssec:coqblin-schrieffer} and \ref{ssec:high-field}.

\subsection*{Acknowledgements}
We thank Leonid Pourovskii, Jernej Mravlje, Houssam El Mrabet Haje, and Kathryn Pitton for fruitful discussions.
We acknowledge the support of HFML-FELIX, a member of the European Magnetic Field Laboratory (EMFL).
We acknowledge the MAX IV Laboratory for beamtime on the Bloch beamline under proposal 20250269.
Research conducted at MAX IV, a Swedish national user facility, is supported by Vetenskapsr\aa{}det (Swedish Research Council, VR) under contract 2018-07152, Vinnova (Swedish Governmental Agency for Innovation Systems) under contract 2018-04969 and Formas under contract 2019-02496.
M.P.D., Z.C., P.M., D.P., A.G.-\v{C}., M.R., S.B., M.E.K., S.W. and M.N.A. acknowledge the research program ``Materials for the Quantum Age'' (QuMat) for financial support.
This program (registration number 024.005.006) is part of the Gravitation program financed by the Dutch Ministry of Education, Culture and Science (OCW).
M.N.A. acknowledges support from the NWO Talent Programme VIDI financed by the NWO VI.Vidi.223.089.
T.M.K. and M.N.A. acknowledges the Kavli Institute Innovation Award 2023.
M.P.D., T.M.K., V.N.V., and M.N.A. acknowledges the Kavli Institute of Nanoscience Delft.
R.M. and J.Y.C. acknowledge NSF DMR-2505304 and Welch Foundation AA-2056-20240404 for partial support of this project.
V.N.V. acknowledges financial support from the HFML-FELIX Incentive fund.
X.L. acknowledges the Guangzhou Elite Project scholarship (S. J. [2022] No. 1).
A.G.-\v{C}. acknowledges the financial support of the Zernike Institute for Advanced Materials.
S.G. and M.R. acknowledge support from the Vidi ENW research programme of the Dutch Research Council (NWO) [Grant DOI: 10.61686/YDRHT18202] with File No. VI.Vidi.233.077.

\subsection*{Data availability}
The data supporting the findings of this study are openly available in 4TU.ResearchData at https://doi.org/10.4121/8807a9d6-8e8d-4484-8e0d-6cd972e61e0e

\subsection*{Code availability}
The code supporting the findings and creating the figures of this study are openly available in 4TU.ResearchData at https://doi.org/10.4121/4ca95cec-f0b2-41e5-81af-3f6b462cd3de

\subsection*{Use of AI tools}
Claude (Claude Opus 5.5, Anthropic) and Codex (GPT-6.1-sol, OpenAI) were used for language editing and refinement of the manuscript text, for refinement of the code used to generate figures, and for refinement of analysis code.
All AI-assisted text and code were reviewed, tested and verified by the authors, and all figures were checked against the underlying data.
The study design, analyses, interpretations and conclusions are the authors' own, and the authors take full responsibility for the content of this article.

\printbibliography

\clearpage
\makeatletter
\long\def\protected@write#1#2#3{\begingroup#2\let\protect\@unexpandable@protect\edef\reserved@a{\immediate\write#1{#3}}\reserved@a\endgroup}
\makeatother
\AddToHook{shipout/before}{\DiscardShipoutBox}
\begin{refsection}
\input{si-body}
\printbibliography
\end{refsection}
\end{document}

%% file: venue-naturemat.tex
\documentclass[11pt]{article}

\usepackage{authblk}

\newcommand{\venueloaded}{naturemat}

\usepackage[letterpaper, top=1in, bottom=1in, left=1in, right=1in, headsep=0.25in]{geometry}

\newlength{\figwidth}
\newlength{\panelwidth}
\AtBeginDocument{%
  \setlength{\figwidth}{\textwidth}%
  \setlength{\panelwidth}{0.48\textwidth}%
}

\PassOptionsToPackage{style=nature}{biblatex}

\PassOptionsToPackage{font=small, labelfont=bf}{caption}

%% file: preamble.tex
\ifdefined\venueloaded\else
  \PackageError{preamble}{No venue file loaded}%
    {Input a venue-*.tex file before preamble.tex.}%
\fi
\ifdefined\figwidth\else
  \PackageError{preamble}{The venue file defines no \string\figwidth}%
    {A venue file must declare the lengths \string\figwidth\space and \string\panelwidth.}%
\fi
\ifdefined\panelwidth\else
  \PackageError{preamble}{The venue file defines no \string\panelwidth}%
    {A venue file must declare the lengths \string\figwidth\space and \string\panelwidth.}%
\fi

\usepackage{graphicx}
\usepackage{subcaption}
\usepackage{float}

\usepackage[section]{placeins}
\makeatletter

\def\fps@figure{htbp}
\def\fps@table{htbp}
\makeatother

\usepackage[version=4]{mhchem}

\input{crystals}

\usepackage{biblatex}
\AtEveryBibitem{%
  \clearfield{urlyear}%
  \clearfield{urlmonth}%
  \clearfield{urlday}%
  \clearfield{issn}%
  \clearfield{month}%
}
\csletcs{blx@bbx@misc@generic}{blx@bbx@misc}
\DeclareFieldFormat{preprint:title}{#1}
\DeclareFieldFormat{preprint:arxiv}{Preprint at \url{https://arxiv.org/abs/#1}}
\DeclareBibliographyDriver{misc}{%
  \iffieldequalstr{eprinttype}{arXiv}
    {\usebibmacro{bibindex}%
     \usebibmacro{begentry}%
     \usebibmacro{author/editor+others/translator+others}%
     \setunit{\labelnamepunct}\newblock
     \printfield[preprint:title]{title}%
     \newunit\newblock
     \printfield[preprint:arxiv]{eprint}%
     \setunit{\addspace}%
     \printtext[parens]{\usebibmacro{date}}%
     \usebibmacro{finentry}}
    {\csuse{blx@bbx@misc@generic}}%
}
\AtBeginBibliography{\emergencystretch=3em}
\usepackage{microtype}
\AtBeginDocument{\microtypesetup{protrusion=false, expansion=false}}
\AtBeginBibliography{\microtypesetup{protrusion=true, expansion=true}\hfuzz=2pt}

\newcommand{\todo}[1]{}
\usepackage{xr-hyper}
\usepackage[hidelinks]{hyperref}

\ExplSyntaxOn
\cs_new:Npn \__paper_ce_pdf_char:n #1
  {
    \str_case_e:nnF {#1}
      { {_} {} {^} {} { \c_left_brace_str } {} { \c_right_brace_str } {} }
      {#1}
  }
\ExplSyntaxOff

%% file: crystals.tex
\makeatletter
\DeclareRobustCommand{\crystal}[2]{%
  \ifcsname crystal@#1@#2\endcsname
    \csname crystal@#1@#2\endcsname
  \else
    \PackageError{crystals}{No value `#2' for crystal `#1'}%
      {Run scripts/write_crystals_tex.py in Pjieter/YbCoGe and copy crystals.tex over.}%
  \fi}
\@namedef{crystal@RM154@formula}{\ce{Yb_{0.48}Co3Ge_{2.95}Sn_{0.05}}}
\@namedef{crystal@RM154@formulaerrors}{\ce{Yb_{0.4758(11)}Co3Ge_{2.951(2)}Sn_{0.049(2)}}}
\@namedef{crystal@RM154@short}{\ce{Yb_{0.48}}}
\@namedef{crystal@RM154@y}{0.48}
\@namedef{crystal@RM154@a}{5.0837(9)}
\@namedef{crystal@RM154@c}{3.8933(9)}
\@namedef{crystal@RM154@split}{1.22}
\@namedef{crystal@RM154@occupancy}{47.6}
\@namedef{crystal@RM154@sn}{4.9}
\@namedef{crystal@RM139@formula}{\ce{Yb_{0.48}Co3Ge_{2.97}Sn_{0.03}}}
\@namedef{crystal@RM139@formulaerrors}{\ce{Yb_{0.4834(9)}Co3Ge_{2.967(2)}Sn_{0.033(2)}}}
\@namedef{crystal@RM139@short}{\ce{Yb_{0.48}}}
\@namedef{crystal@RM139@y}{0.48}
\@namedef{crystal@RM139@a}{5.0871(9)}
\@namedef{crystal@RM139@c}{3.9150(9)}
\@namedef{crystal@RM139@split}{1.22}
\@namedef{crystal@RM139@occupancy}{48.3}
\@namedef{crystal@RM139@sn}{3.3}
\@namedef{crystal@RM230@formula}{\ce{Yb_{0.37}Co3Ge_{2.74}Sn_{0.26}}}
\@namedef{crystal@RM230@formulaerrors}{\ce{Yb_{0.3681(12)}Co3Ge_{2.736(2)}Sn_{0.264(2)}}}
\@namedef{crystal@RM230@short}{\ce{Yb_{0.37}}}
\@namedef{crystal@RM230@y}{0.37}
\@namedef{crystal@RM230@a}{5.1058(9)}
\@namedef{crystal@RM230@c}{3.8937(7)}
\@namedef{crystal@RM230@split}{1.21}
\@namedef{crystal@RM230@occupancy}{36.8}
\@namedef{crystal@RM230@sn}{26.4}
\@namedef{crystal@RM048@formula}{\ce{Yb_{0.36}Co3Ge_{2.73}Sn_{0.27}}}
\@namedef{crystal@RM048@formulaerrors}{\ce{Yb_{0.3647(12)}Co3Ge_{2.729(2)}Sn_{0.271(2)}}}
\@namedef{crystal@RM048@short}{\ce{Yb_{0.36}}}
\@namedef{crystal@RM048@y}{0.36}
\@namedef{crystal@RM048@a}{5.1101(10)}
\@namedef{crystal@RM048@c}{3.8899(10)}
\@namedef{crystal@RM048@split}{1.21}
\@namedef{crystal@RM048@occupancy}{36.5}
\@namedef{crystal@RM048@sn}{27.1}
\@namedef{crystal@RM238@formula}{\ce{Yb_{0.19}Co3Ge_{2.38}Sn_{0.62}}}
\@namedef{crystal@RM238@formulaerrors}{\ce{Yb_{0.1888(14)}Co3Ge_{2.378(3)}Sn_{0.623(3)}}}
\@namedef{crystal@RM238@short}{\ce{Yb_{0.19}}}
\@namedef{crystal@RM238@y}{0.19}
\@namedef{crystal@RM238@a}{5.1010(12)}
\@namedef{crystal@RM238@c}{3.8773(13)}
\@namedef{crystal@RM238@split}{1.07}
\@namedef{crystal@RM238@occupancy}{18.9}
\@namedef{crystal@RM238@sn}{62.3}
\@namedef{crystal@RM176@formula}{\ce{Y_{0.22}Co3Ge_{2.45}Sn_{0.55}}}
\@namedef{crystal@RM176@formulaerrors}{\ce{Y_{0.2241(13)}Co3Ge_{2.448(2)}Sn_{0.552(2)}}}
\@namedef{crystal@RM176@short}{\ce{Y_{0.22}}}
\@namedef{crystal@RM176@y}{0.22}
\@namedef{crystal@RM176@a}{5.1181(11)}
\@namedef{crystal@RM176@c}{3.8787(10)}
\@namedef{crystal@RM176@split}{1.14}
\@namedef{crystal@RM176@occupancy}{22.4}
\@namedef{crystal@RM176@sn}{55.2}
\@namedef{crystal@NJF043@formula}{\ce{Tb_{0.34}Co3Ge_{2.66}Sn_{0.34}}}
\@namedef{crystal@NJF043@formulaerrors}{\ce{Tb_{0.338(3)}Co3Ge_{2.656(6)}Sn_{0.344(5)}}}
\@namedef{crystal@NJF043@short}{\ce{Tb_{0.34}}}
\@namedef{crystal@NJF043@y}{0.34}
\@namedef{crystal@NJF043@a}{5.123(2)}
\@namedef{crystal@NJF043@c}{3.903(2)}
\@namedef{crystal@NJF043@split}{1.19}
\@namedef{crystal@NJF043@occupancy}{33.8}
\@namedef{crystal@NJF043@sn}{32.4}
\makeatother

%% file: authors.tex
\author[1,2,*]{M.~P.~Dubbelman}
\author[3]{R.~Meduri}
\author[1,2]{T.~M.~Kyrk}
\author[4]{Z.~Cheng}
\author[5]{P.~Maderson}
\author[6]{D.~Pizzirani}
\author[8]{A.~Das}
\author[1,2,4,6]{V.~N.~Vreede}
\author[9]{X.~Liu}
\author[10]{N.~Witt}
\author[7]{S.~Grytsiuk}
\author[11]{C.~M.~Polley}
\author[10]{M.~R\"{o}sner}
\author[9]{A.~Grubi\v{s}i\'{c}-\v{C}abo}
\author[4]{S.~Bhattacharyya}
\author[8]{R.~E.~Baumbach}
\author[5]{M.~E.~Kamminga}
\author[6,7]{S.~Wiedmann}
\author[3]{J.~Y.~Chan}
\author[1,2,**]{M.~N.~Ali}

\affil[1]{Department of Quantum Nanoscience, Delft University of Technology, 2628 CJ Delft, the Netherlands}
\affil[2]{Kavli Institute of Nanoscience, Delft University of Technology, 2628 CJ Delft, the Netherlands}
\affil[3]{Department of Chemistry and Biochemistry, Baylor University,
Waco, TX 76798, USA}
\affil[4]{Leiden Institute of Physics, Leiden University, 2333 CA Leiden, the Netherlands}
\affil[5]{Debye Institute for Nanomaterials Science, Utrecht University, 3508 TA Utrecht, the Netherlands}
\affil[6]{HFML-FELIX, 6525 ED Nijmegen, the Netherlands}
\affil[7]{Institute for Molecules and Materials, Radboud University, 6525 AJ Nijmegen, the Netherlands}
\affil[8]{Department of Physics, University of California, Santa Cruz,
Santa Cruz, CA 95064, USA}
\affil[9]{Zernike Institute for Advanced Materials, University of Groningen,
9747 AG Groningen, the Netherlands}
\affil[10]{Faculty of Physics, Bielefeld University,
33615 Bielefeld, Germany}
\affil[11]{MAX IV Laboratory, Lund University, Lund 22100, Sweden}
\affil[*]{m.p.dubbelman@tudelft.nl}
\affil[**]{m.n.ali@tudelft.nl}

%% file: si-body.tex

\setcounter{section}{0}
\setcounter{figure}{0}
\setcounter{table}{0}
\setcounter{equation}{0}
\renewcommand{\thefigure}{S\arabic{figure}}
\renewcommand{\thetable}{S\arabic{table}}
\renewcommand{\theequation}{S\arabic{equation}}
\renewcommand{\theHsection}{S\arabic{section}}
\renewcommand{\theHfigure}{S\arabic{figure}}
\renewcommand{\theHtable}{S\arabic{table}}
\renewcommand{\theHequation}{S\arabic{equation}}

\section{Samples}
\label{ssec:samples}

Supplementary Table~\ref{stab:inventory} lists the crystals of this work and what was measured on each.
The suffixes .1 and .3 tell apart the crystals of one growth batch described in the Methods of the main text.

\begin{figure}
    \centering
    \includegraphics[width=\figwidth]{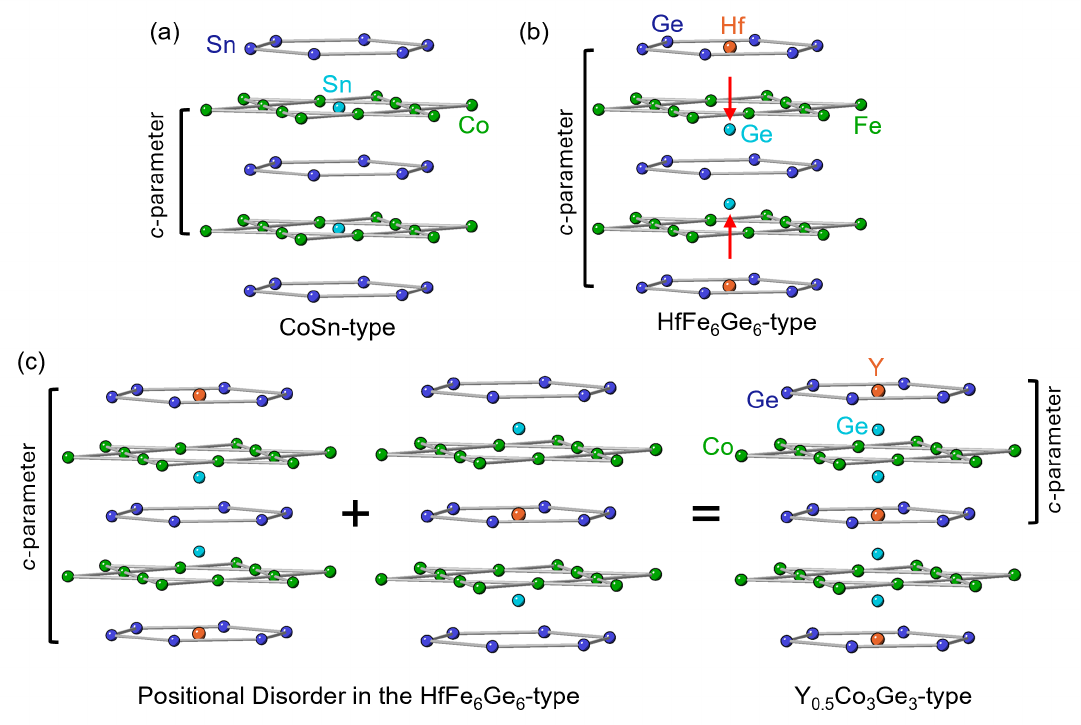}
    \caption{Derivation of the \ce{Y_{0.5}Co3Ge3} structure type.
    (a) \ce{CoSn} type, with honeycomb slabs of the group-14 element alternating with Co kagome slabs centered by the same element.
    (b) \ce{HfFe6Ge6} type.
    An electropositive atom in every other honeycomb layer pushes the kagome-centered atoms towards the neighboring honeycomb layer (red arrows), which doubles $c$.
    (c) Disorder between the two equivalent placements of this pattern, 50~\% each, averages to the \ce{Y_{0.5}Co3Ge3} type with the original $c$.}
    \label{sfig:structure-derivation}
\end{figure}

The \ce{Yb_{0.5}Co3Ge3} structure type \autocite{Weiland_etal_2020} is a disordered member of the \ce{AM6X6} family.
The single-crystal refinement of \crystal{RM154}{formula} finds the Yb site \crystal{RM154}{occupancy}~\% occupied and a split Ge site \crystal{RM154}{split}~\AA{} from it, which is too close for both to be filled in the same cell, so each cell holds either a Yb atom or a Ge pair across the empty Yb site.
Together with \crystal{RM154}{sn}~\% Sn on the kagome-centered site, this decomposes the crystal into \crystal{RM154}{occupancy}~\% of each of the two \ce{Y_{0.5}Co3Ge3}-type configurations and \crystal{RM154}{sn}~\% of the \ce{CoSn} type (main-text Fig.~1b), consistent with the hybrid structure reported previously \autocite{Weiland_etal_2020}.
Supplementary Fig.~\ref{sfig:structure-derivation} derives the \ce{Y_{0.5}Co3Ge3} type from the \ce{CoSn} type.

\begin{table}
    \centering
    \small
    \begin{tabular}{llp{6.4cm}l}
        \hline
        crystal & filling & measurements & main-text figures \\
        \hline
        RM154.1 & \crystal{RM154}{short} & $\rho(T)$ at 0 to 5~T, and at 9 to 30~T on V56; magnetoresistance to 32~T for $B \perp c$ and $B \parallel c$ on contacts V56, and for $B \parallel c$ on V12 & 1d, 2, 3a \\
        RM154.3 & \crystal{RM154}{short} & $M(T)$ from 1.8 to 300~K and $M(B)$ from 2 to 30~K up to 5~T, both orientations & 2 (insets) \\
        RM088 & \crystal{RM154}{short} & ARPES at 18~K with $h\nu = 188$~eV, along $\Gamma$--K--M & 4a, b \\
        RM139.1 & \crystal{RM139}{short} & $C(T)$ at 0 to 9~T for $B \perp c$, from 2.2 to 30~K, and to 200~K at zero field & 1e \\
        RM230.1 & \crystal{RM230}{short} & $\rho(T)$ at 0 to 5~T, and at 9 to 30~T on V34; magnetoresistance to 32~T for $B \parallel c$ and $B \perp c$ on contacts V34 & 1d, 3b \\
        RM230.3 & \crystal{RM230}{short} & $M(T)$ from 1.8 to 300~K and $M(B)$ from 2 to 30~K up to 5~T, both orientations & \\
        RM048.1 & \crystal{RM048}{short} & $C(T)$ at 0 to 9~T for $B \perp c$, from 2.2 to 20~K, and to 120~K at zero field & 1e \\
        RM238.1 & \crystal{RM238}{short} & $\rho(T)$ at 0 to 5~T & 1d \\
        RM176.1 & \crystal{RM176}{short} & $\rho(T)$, Y analog & 1d \\
        NJF043.1 & \crystal{NJF043}{short} & $\rho(T)$, Tb analog & 1d \\
        \hline
    \end{tabular}
    \caption{Crystals of this work and the measurements done on them.
    The filling is the rare-earth count of the refinement of each batch.
    RM088 has the composition of RM154.
    In every transport crystal the current runs along $c$.}
    \label{stab:inventory}
\end{table}

\begin{table}
    \centering
    \small
    \begin{tabular}{llccl}
        \hline
        batch & refined formula & $a$ (\AA) & $c$ (\AA) & melt $R$:Co:Ge:Sn \\
        \hline
        RM154 & \crystal{RM154}{formulaerrors} & \crystal{RM154}{a} & \crystal{RM154}{c} & 3:2:7:52 \\
        RM139 & \crystal{RM139}{formulaerrors} & \crystal{RM139}{a} & \crystal{RM139}{c} & 3:2:7:20 \\
        RM230 & \crystal{RM230}{formulaerrors} & \crystal{RM230}{a} & \crystal{RM230}{c} & 1:6:6:25 \\
        RM048 & \crystal{RM048}{formulaerrors} & \crystal{RM048}{a} & \crystal{RM048}{c} & 1.2:6:6:25 \\
        RM238 & \crystal{RM238}{formulaerrors} & \crystal{RM238}{a} & \crystal{RM238}{c} & 1:6:6:10 \\
        RM176 & \crystal{RM176}{formulaerrors} & \crystal{RM176}{a} & \crystal{RM176}{c} & 1:6:6:20 \\
        NJF043 & \crystal{NJF043}{formulaerrors} & \crystal{NJF043}{a} & \crystal{NJF043}{c} & 1.2:6:6:20 \\
        \hline
    \end{tabular}
    \caption{Growth and characterization of the seven crystals.
    All Yb batches were grown with the temperature profile of RM154 in the Methods of the main text, and the Y and Tb analogs with their own profiles given there.
    The melt ratio is the molar loading, with $R$ the rare earth.
    Formulas and lattice parameters are from the single-crystal refinements at room temperature, with the standard uncertainty of the last digits in brackets.}
    \label{stab:growth}
\end{table}

\section{Resistance in field}
\label{ssec:resistance-in-field}

\begin{figure}
    \centering
    \includegraphics[width=\figwidth]{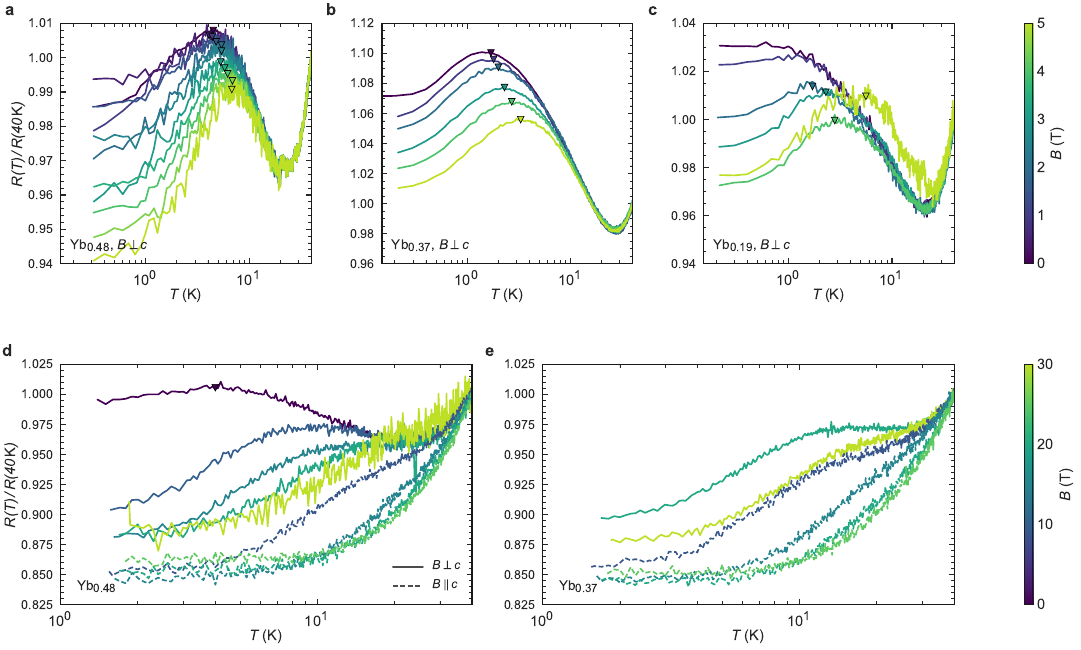}
    \caption{Resistance against temperature at every measured field, normalized to its value at 40~K.
    (a--c) $B \perp c$ and $I \parallel c$ from 0 to 5~T, on the contacts of main-text Fig.~1d: \crystal{RM154}{formula} in steps of 0.5~T, \crystal{RM230}{formula} and \crystal{RM238}{formula} in steps of 1~T.
    Triangles mark the coherence maxima, found as in main-text Fig.~1d.
    (d, e) Fixed fields from 9 to 30~T in the magnet of main-text Fig.~3, on the same contacts, for \crystal{RM154}{formula} and \crystal{RM230}{formula}, with $B \perp c$ solid and $B \parallel c$ dashed.
    Each sweep starts at the 1.35~K bath, where the thermometer reads up to 0.5~K high in field from its own magnetoresistance; temperatures are shown as read.
    Only warming sweeps are used in (c).
    The 5~T curve in (c) has a lock-in phase that drifts during the sweep and lies about 2~\% above the other fields between 3 and 20~K.}
    \label{sfig:resistance-in-field}
\end{figure}

Main-text Fig.~1d is measured at 1~T, which suppresses the superconductivity of Sn inclusions from the flux \autocite{Wang_etal_2022}.
Supplementary Fig.~\ref{sfig:resistance-in-field} shows the resistance of the three Yb crystals against temperature at every field it was measured in, the fields of the inset of main-text Fig.~1d included.

\section{Heat capacity}
\label{ssec:heat-capacity}

\subsection{Resonant-level heat capacity}
\label{ssec:resonant-level}

The heat capacity per mole of Yb is fitted with
\begin{equation}
    C = \gamma T + \beta T^3 + n R\, C_\mathrm{res}(T; \Gamma, \Delta),
\end{equation}
where $\gamma$ is the Sommerfeld coefficient of the conduction electrons, $\beta T^3$ the phonon term, $n$ the number of Kondo-broadened Kramers doublets per Yb and $R$ the gas constant. 
In the model of Schotte and Schotte \autocite{Schotte_Schotte_1975} the Kondo-screened doublet acts as a single resonant level at the Zeeman energy $\Delta = g_\perp \mu_\mathrm{B} B / k_\mathrm{B}$, broadened into a Lorentzian with half-width $\Gamma$, both in kelvin.
Its heat capacity per doublet, in units of $R$, is the two-level Schottky term averaged over that Lorentzian,
\begin{equation}
    C_\mathrm{res} = \int_{-\infty}^{\infty} \frac{\Gamma/\pi}{(\varepsilon - \Delta)^2 + \Gamma^2} \left(\frac{\varepsilon}{2T}\right)^2 \mathrm{sech}^2\frac{\varepsilon}{2T}\, d\varepsilon = \mathrm{Re}\left[2z\left(1 - z\,\psi'(1/2 + z)\right)\right], \; z = \frac{\Gamma - i\Delta}{2\pi T},
\end{equation}
with $\psi'$ the trigamma function.
For $\Gamma \to 0$ it reduces to the Schottky anomaly of a doublet split by $\Delta$.
At zero field and $T \ll \Gamma$ it is linear, $C_\mathrm{res} \to \pi T/(3\Gamma)$, so the doublets add to the Sommerfeld coefficient and
\begin{equation}
    \gamma(0) = \gamma + \frac{n R \pi}{3\Gamma}
\end{equation}
is the zero-field $C/T$ at $T \to 0$ reported per mole of Yb.

The parameters $\gamma$, $n$, $\Gamma$ and $g_\perp$ are fitted to all fields from 0 to 9~T at once, below 6~K, by least squares on the relative residual $(C_\mathrm{fit} - C)/C$.
$\beta$ is held at the low-temperature limit of a Debye plus Einstein lattice fit to the zero-field data from 25~K up to the highest measured temperature (Methods of the main text), $\beta = 12\pi^4 N_\mathrm{D} R/(5\theta_\mathrm{D}^3 y)$, with $\theta_\mathrm{D}$ the Debye temperature, $N_\mathrm{D}$ the number of atoms per formula unit in the Debye branch, and $y$ the number of Yb atoms per formula unit to normalize it again.

From the fit we extract $\Gamma = 4.4$~K and $g_\perp = 3.96$, with $n = 0.85$ for \crystal{RM139}{short} and $\Gamma = 3.3$~K and $g_\perp = 1.69$, with $n=0.74$ for \crystal{RM048}{short}.
$\gamma(0)$ is an extrapolation below the lowest measured temperature, but drawing on all ten fields at once makes it more robust than a fit to a single field.
The two crystals, with different Yb contents, give $\gamma(0)$ within 15~\% of each other per Yb ion, implying that the mass enhancement is a property of the Yb ion in this host and not of the specific filling.

\subsection{4f entropy and phonon backgrounds}
\label{ssec:entropy}

\begin{figure}[H]
    \centering
    \includegraphics[width=\panelwidth]{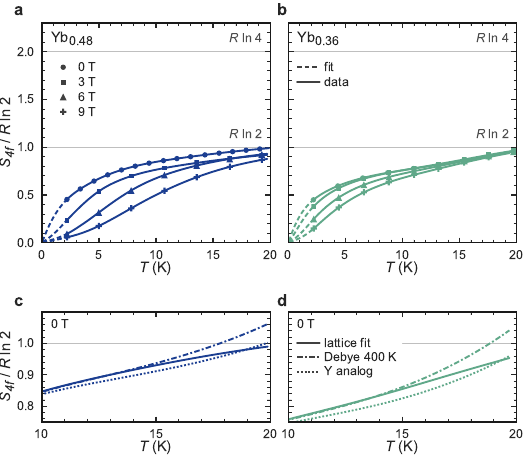}
    \caption{4f entropy per mole Yb of (a, c) \crystal{RM139}{formula} and (b, d) \crystal{RM048}{formula}, with $B \perp c$.
    (a, b) 0, 3, 6 and 9~T.
    Below the lowest measured temperature, near 2.2~K, the entropy is that of the resonant-level fit of main-text Fig.~1e, drawn dashed.
    Above it the measured heat capacity less the lattice-fit phonon background is integrated, drawn solid.
    The gray lines mark the entropy of one Kramers doublet, $R \ln 2$, and of two, $R \ln 4$.
    (c, d) 0~T between 10 and 20~K with each of the three phonon backgrounds: the lattice fit, a 400~K Debye background and the cubic lattice term of the Y analog.
    Section~\ref{ssec:entropy} defines the integral and the three backgrounds.}
    \label{sfig:entropy}
\end{figure}

The resonant-level fit has no crystal-field term.
To check that no excited crystal-field level is populated below 20~K, we count the 4f entropy per mole Yb (Supplementary Fig.~\ref{sfig:entropy}),
\begin{equation}
    S_{4f}(T) = \frac{1}{y}\left[\int_0^{T_0} \frac{\gamma T' + n R\, C_\mathrm{res}(T')}{T'}\, dT' + \int_{T_0}^{T} \frac{C(T') - C_\mathrm{ph}(T')}{T'}\, dT'\right],
\end{equation}
where $y$ is the Yb filling and $T_0 \approx 2.2$~K the lowest measured temperature.
Below $T_0$ the integrand is the resonant-level fit without its phonon term, and above $T_0$ the measured heat capacity minus a phonon background $C_\mathrm{ph}$.
Our data cannot separate the lattice from the 4f part below 20~K, so we compare three backgrounds: the Debye plus Einstein lattice fit that sets $\beta$, a single Debye branch with $\theta_\mathrm{D} = 400$~K, and the cubic term $\beta = 0.78$~mJ\,mol$_{\mathrm{Y}}^{-1}$\,K$^{-4}$ of the Y analog \autocite{Bravo_etal_2023}.
At 20~K they make up 78 to 84~\%, 41 to 46~\% and 55 to 61~\% of the measured heat capacity.
Over all three backgrounds and all fields from 0 to 9~T, $S_{4f}$ at 20~K lies between 0.88 and $1.07\,R\ln 2$ in both crystals.
This is the entropy of the ground Kramers doublet alone, far below the $R\ln 4$ of two populated doublets, so the fit needs no crystal-field term.
The conduction electrons add approximately 0.07 to $0.09\,R\ln 2$ to this, taking the Sommerfeld coefficient of the Y analog \autocite{Bravo_etal_2023}, so $S_{4f}$ is an upper limit, and moving $T_0$ to 4 or 6~K changes it by at most $0.03\,R\ln 2$.

\section{X-ray photoemission}
\label{ssec:xps}

X-ray photoemission spectra were collected with a Surface Science Instruments SSX-100 ESCA spectrometer with a monochromatic Al K$\alpha$ source ($h\nu = 1486.6$~eV).
The electron take-off angle was 37$^\circ$ from the surface normal.
The samples were mounted on carbon tape.
Every spectrum was taken with a spot 1000~$\mu$m in diameter and an energy resolution of 1.3~eV.
The analyzer was referenced to the Au 4f peak at a binding energy of 84.0~eV, and binding energies carry an uncertainty of $\pm 0.2$~eV.
The spectra were fitted by least squares in WinSpec, developed at LISE, University of Namur, Belgium, with a Shirley background and peaks that are a linear combination of a Gaussian and a Lorentzian, taking the experimental resolution into account.
Each spectrum was fitted with the smallest number of peaks consistent with the structure of the surface.

The Yb 4d and Ge 3s region of \crystal{RM154}{formula} and \crystal{RM048}{formula} are shown in Supplementary Fig.~\ref{sfig:xps}.
Both spectra show the broad multiplet of \ce{Yb^3+}.
Strong Coulomb and exchange interactions in the 4d final state split it into components spread over approximately 20~eV.
The largest lies at a binding energy of approximately 186.6~eV, with further features at approximately 189.0, 194.5, 200.3 and 205.6~eV.
The shape of the multiplet and the relative peak positions agree with those reported for \ce{Yb^3+} compounds \autocite{Ohno_2008,Schmidt_etal_2005,Chorkendorff_etal_1985}.
The Yb is therefore predominantly trivalent in both crystals, in agreement with the magnetization.

We assign the features at approximately 181.2 and 184.6~eV to metallic Ge (\ce{Ge^0}) and Ge oxide (\ce{GeO_x}).
In \crystal{RM048}{short} a further component near 200.5~eV is Cl 2p, from residual \ce{SnCl2} and C--Cl species.
The Ge and Cl components cannot be quantified independently within the Yb 4d envelope, so we identified and quantified them in the Ge 3p and Cl 2s spectra, which do not overlap it.
Their areas, corrected with the relative sensitivity factors, enter the fits of the Yb 4d region as fixed components.

\begin{figure}
    \centering
    \includegraphics[width=\figwidth]{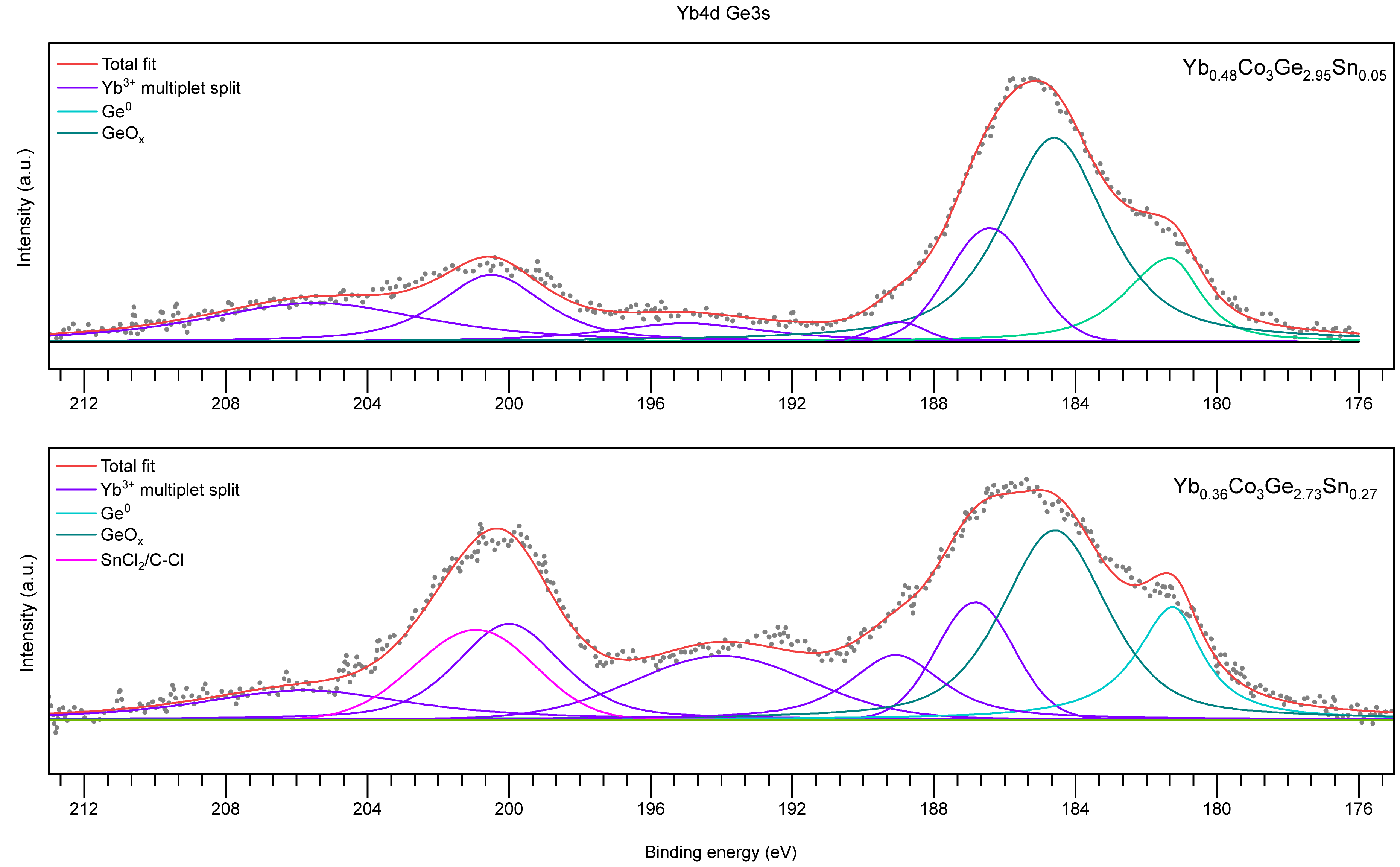}
    \caption{Yb 4d and Ge 3s X-ray photoemission spectra of (top) \crystal{RM154}{formula} and (bottom) \crystal{RM048}{formula}.
    Gray dots are data and red lines the total fit, the sum of the \ce{Yb^3+} multiplet (purple), metallic Ge (cyan), Ge oxide (dark cyan) and, bottom only, \ce{SnCl2} or C--Cl (magenta).
    Both spectra are fitted with \ce{Yb^3+} alone.
    Section~\ref{ssec:xps} describes the fits.}
    \label{sfig:xps}
\end{figure}

\section{Single-ion Kondo scaling}
\label{ssec:scaling}

\subsection{Magnetoresistance conventions}
\label{ssec:mr-conventions}

Each resistivity isotherm is symmetrized, $\rho(B) = [\rho(+B) + \rho(-B)]/2$, and binned in 0.25~T bins, taking the median of each bin.
The zero-field resistivity $\rho_0$ in $\Delta\rho/\rho_0$ is the mean of the bins below 0.5~T.

\subsection{Standard error}
\label{ssec:standard-error}

The standard error of a fit is $\sqrt{\sum r^2/(N_B - p)}$, with $r$ the unweighted residuals in percentage points (pp), $N_B$ the number of field points and $p$ the number of free parameters.

\subsection{Schlottmann magnetization and resistivity}
\label{ssec:coqblin-schrieffer}

The Schlottmann model is the Bethe-ansatz solution of the Coqblin-Schrieffer model,\autocite{Schlottmann_1983} valid for independent single-ion moments.
For $j = 1/2$ and $T = 0$ it gives the magnetization of a Kondo-screened ion, in units of the saturated moment $j$, as a function of $t = \ln(B/B^*)$ alone,
\begin{equation}
    M(t) = \frac{1}{4} - \frac{1}{2\pi^{3/2}} \int_0^\infty \frac{du}{u}\, \mathrm{Im}\left[\Gamma\!\left(\frac{1}{2} + \frac{iu}{2}\right) e^{-iut} (-iu)^{-iu/2}\right],
\end{equation}
with $\Gamma(x)$ Euler's gamma function.
At low field $M \to (B/B^*)/\sqrt{\pi}$, so $B^*$ sets the initial susceptibility.
The Friedel sum rule \autocite{Schlottmann_1983} turns the occupations $1/2 \pm M$ of the doublet into the resistivity of the ion relative to zero field,
\begin{equation}
    \mathcal{R}(t) = \cos^2\left[\pi M(t)\right],
\end{equation}
which falls from 1 to 0 at full polarization.
$\mathcal{R}(t)$ is the fit line in main-text Fig.~2, and $M(t)$ the fit line in its insets.

\begin{figure}
    \centering
    \includegraphics[width=\figwidth]{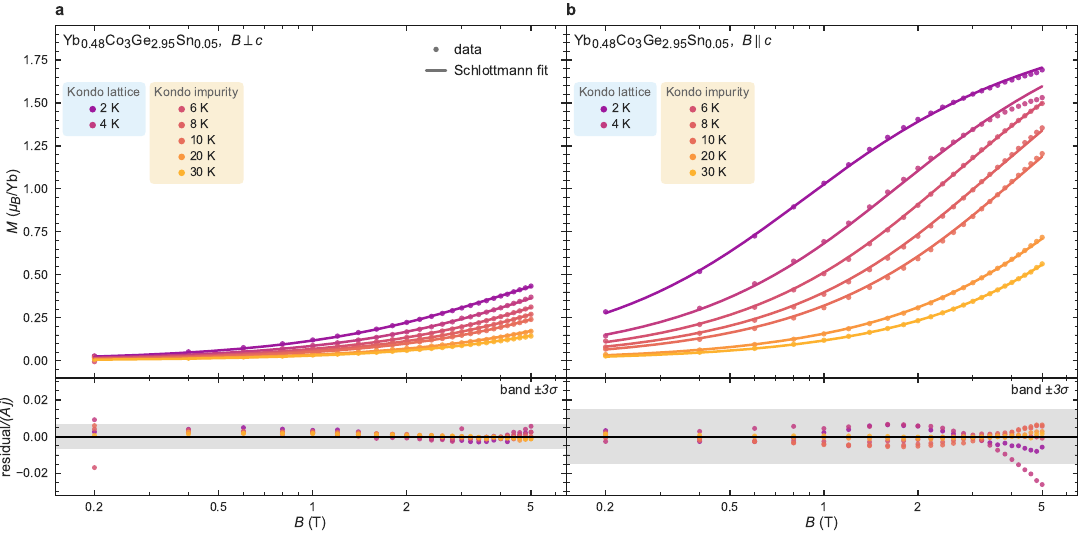}
    \todo{The magnetoresistance residuals of RM230.1, which the original plan (issue 36) had beside RM154.1, are not in the paper yet.}
    \caption{Magnetization isotherms between 2 and 30~K of a crystal from the same batch as \crystal{RM154}{formula}, behind the insets of main-text Fig.~2.
    (a) $B \perp c$.
    (b) $B \parallel c$.
    Markers are data and lines the joint Schlottmann magnetization fit of each orientation.
    The lower panels show the residuals in units of $Aj$.
    Gray bands span $\pm 3\sigma$, with $\sigma$ the root-mean-square residual, 0.0023 for $B \perp c$ and 0.0049 for $B \parallel c$.}
    \label{sfig:mh-isotherms}
\end{figure}

The magnetoresistance isotherms of one orientation are fitted at once with
\begin{equation}
    \frac{\Delta\rho}{\rho_0}(B, T_i) = A_i \left[\mathcal{R}\!\left(\ln\frac{B}{B^*(T_i)}\right) - 1\right], \qquad B^*(T) = \sqrt{B_0^2 + (sT)^2},
\end{equation}
where $A_i$ is the fraction of $\rho_0$ that full polarization removes, $B_0$ the zero-temperature field scale, and $T^* = B_0/s$ the crossover to thermal broadening.
The magnetization isotherms are fitted in the same way, $M_i = A_i M(\ln[B/B^*_M(T_i)])$, with their own $B_0$ and $s$ (Supplementary Fig.~\ref{sfig:mh-isotherms}).
Both fits are weighted by their inverse measurement error, and isotherms that return $A_i \le 0$ are dropped.
For $B \perp c$ the magnetization reaches at most $B/B^*_M \approx 0.55$ at 5~T, where amplitude and field scale still trade off, so its errors are large.

The $B \parallel c$ magnetoresistance residuals have structure.
At and below 4.3~K they undershoot the model at low field, possibly linked to the onset of coherence, and overshoot it around $B \approx B^*$.
Beyond the polarization of the doublet all isotherms fall below the model.

\begin{figure}
    \centering
    \includegraphics[width=\figwidth]{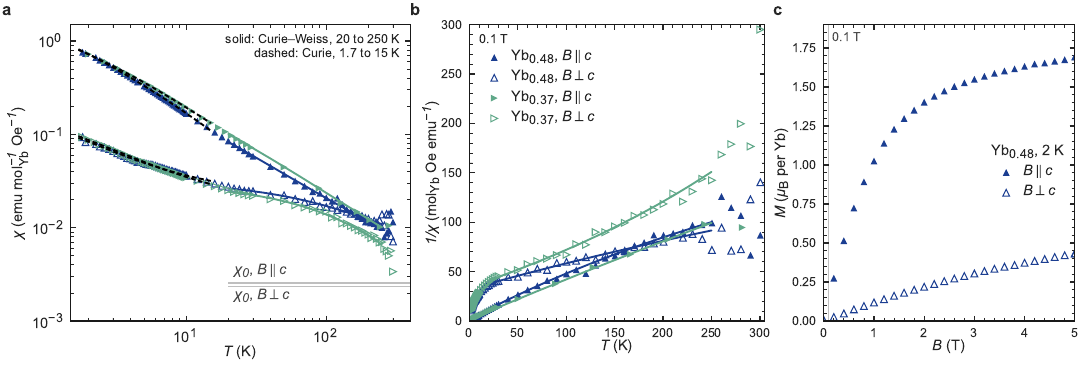}
    \caption{Magnetic susceptibility per mole Yb, with an easy axis along $c$.
    Blue upward triangles are a data points of the \crystal{RM154}{formula} batch, green right-pointing triangles the \crystal{RM230}{formula} batch.
    Filled markers are $B \parallel c$, open markers $B \perp c$.
    (a) Susceptibility at 0.1~T.
    The two batches nearly coincide below 20~K, and at 1.7~K the susceptibility along $c$ is about eight times the in-plane value.
    Colored lines are Curie--Weiss fits with a constant term $\chi_0$ from 20 to 250~K, and black dashed lines are Curie fits from 1.7 to 15~K that give the $g$ factors of Section~\ref{ssec:g-factors}.
    Gray lines mark $\chi_0$ of the \crystal{RM154}{short} crystal, $2.6(3) \times 10^{-3}$ along $c$ and $2.4(12) \times 10^{-3}$~emu\,mol$_\mathrm{Yb}^{-1}$\,Oe$^{-1}$ in the plane.
    (b) Inverse susceptibility with the same fits, which gives Weiss temperatures of $-3$ to $-4$~K, with $B\parallel c$ and $-107$ to $-115$~K with $B\perp c$.
    Above 250~K the signal approaches the background, increasing the scatter.
    (c) Magnetization of the \crystal{RM154}{short} crystal at 2~K, on the sweep down from 5~T.
    The gray line marks 0.1~T, where both orientations are still linear in field.}
    \label{sfig:curie-weiss}
\end{figure}

\section{High-field longitudinal magnetoresistance}
\label{ssec:high-field}

\subsection{Kondo resistivity with an extra channel}
\label{ssec:extra-channel}

In main-text Fig.~3 each $B \parallel c$ isotherm is fitted on its own.
The Kondo part is the Schlottmann resistivity
\begin{equation}
    \frac{\rho_\mathrm{Kondo,SM}}{\rho_0} = 1 + A\left[\mathcal{R}\!\left(\ln\frac{B}{B^*}\right) - 1\right],
\end{equation}
with $A$ and $B^*$ free for every isotherm.
An extra channel conducts in parallel with it, with a conductance $(B/B_\mathrm{ch})^2$ relative to the zero-field conductance of the sample, so that
\begin{equation}
    \frac{\rho}{\rho_0} = \left[\frac{\rho_0}{\rho_\mathrm{Kondo,SM}} + \left(\frac{B}{B_\mathrm{ch}}\right)^2\right]^{-1}, \qquad \rho(B^2) = \rho_\mathrm{Kondo,SM} - \rho.
\end{equation}
At $B = B_\mathrm{ch}$ the channel alone conducts as well as the whole sample at zero field, so a lower $B_\mathrm{ch}$ means a stronger channel.
The fit runs from 0.3 to 32~T.
The channel counts as resolved when the error of $B_\mathrm{ch}$ is at most a quarter of $B_\mathrm{ch}$, and the split is undetermined when $B^*$ plus two errors lies above the highest field.
In \crystal{RM154}{formula} $B_\mathrm{ch}$ rises from 110 to 148~T between 0.44 and 10~K.
In \crystal{RM230}{formula} $B_\mathrm{ch}$ is higher at every temperature below 3~K and rises faster, from 149~T at 0.44~K to 338~T at 2.7~K.

\subsection{Weak localization}
\label{ssec:weak-localization}

The three-dimensional weak-localization correction to the conductivity is \autocite{Kawabata_1980}
\begin{equation}
    \Delta\sigma = \frac{e^2}{2\pi^2\hbar}\sqrt{\frac{eB}{\hbar}}\, f_3\!\left(\frac{B_\varphi}{B}\right), \qquad f_3(x) = \sum_{k=0}^{\infty}\left[2\left(\sqrt{k + 1 + x} - \sqrt{k + x}\right) - \frac{1}{\sqrt{k + 1/2 + x}}\right],
\end{equation}
where the dephasing field $B_\varphi = \hbar/(4eL_\varphi^2)$ is set by the phase-coherence length $L_\varphi$.
The correction is largest for $L_\varphi \to \infty$, where $f_3(0) \approx 0.605$.
At its universal amplitude this is the largest negative magnetoresistance weak localization can give, with no free parameter, and strong spin-orbit scattering would turn it positive.
Added in parallel to the measured zero-field resistivity $\rho_0$, in $\Omega$\,m, it gives
\begin{equation}
    \frac{\Delta\rho}{\rho_0} = \frac{1}{1 + \rho_0 \Delta\sigma} - 1.
\end{equation}
In main-text Fig.~3c $B_\varphi = 0$ and nothing is fitted.
In Supplementary Figs.~\ref{sfig:model-comparison} and \ref{sfig:model-comparison-rm230} $\Delta\sigma$ is multiplied by a free factor and $B_\varphi$ is free as well.
The fit then needs 1200 to 2200 times the universal amplitude in \crystal{RM154}{short} and 1900 to 2200 times in \crystal{RM230}{short}, with the dephasing field driven to zero at every isotherm up to 4.3~K, and still leaves 3 to 14 times the standard error of $\rho_\mathrm{Kondo,SM} - \rho(B^2)$.

\subsection{Spin-disorder scattering}
\label{ssec:yosida}

Scattering of the conduction electrons by the local moments \autocite{Yosida_1957,Beal-Monod_Weiner_1968,Ding_etal_2021} gives a negative magnetoresistance proportional to the square of the magnetization, $\Delta\rho/\rho_0 \propto -M^2$.
A free doublet has $M \propto \tanh(g\mu_\mathrm{B}B/2k_\mathrm{B}T)$.
Replacing $T$ by $T + T^*_\mathrm{YS}$, the usual form for a Kondo-screened moment, gives
\begin{equation}
    \frac{\rho_\mathrm{Kondo,YS}}{\rho_0} - 1 = -A\tanh^2\frac{B}{B_1}, \qquad B_1 = \frac{2k_\mathrm{B}(T + T^*_\mathrm{YS})}{g_\parallel\mu_\mathrm{B}},
\end{equation}
with $A$ the magnetoresistance at full polarization.
The moment of a doublet is bounded, so this term saturates within a few $B_1$ whatever its parameters, while the measured magnetoresistance keeps falling up to 32~T.

\begin{figure}
    \centering
    \includegraphics[width=\figwidth]{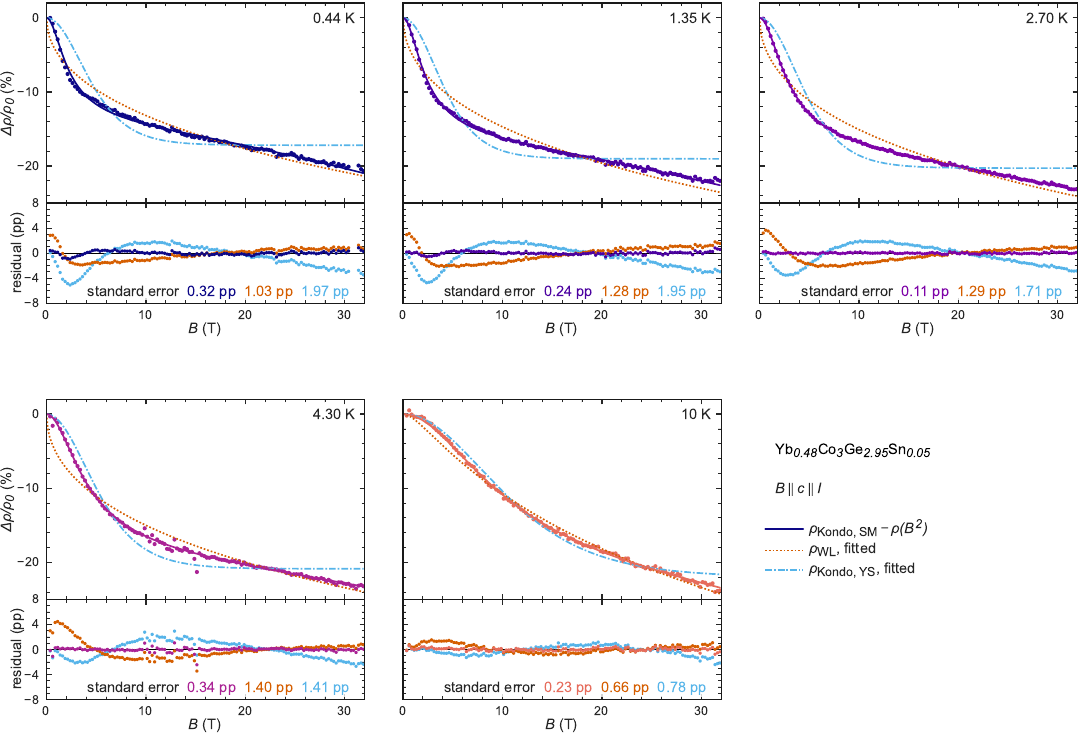}
    \caption{
    Weak localization and Yosida's spin-disorder term fitted freely to every isotherm of main-text Fig.~3a.
    Each isotherm of \crystal{RM154}{formula} is shown with the fit $\rho_\mathrm{Kondo,SM} - \rho(B^2)$ of main-text Fig.~3a, with $\rho_\mathrm{WL}$ \autocite{Kawabata_1980} with free amplitude and dephasing field, and with $\rho_\mathrm{Kondo,YS}$ \autocite{Yosida_1957} (Section~\ref{ssec:yosida}) with free amplitude and field scale $B_1$, each fitted from 0.3 to 32~T without $\rho_\mathrm{Kondo,SM}$.
    Residuals are in percentage points (pp), and the standard errors are as defined in Section~\ref{ssec:standard-error}.}
    \label{sfig:model-comparison}
\end{figure}

In main-text Fig.~3c $B_1$ is fixed by a joint fit of the five isotherms of main-text Fig.~3a from 0.3 to 5~T, with $g_\parallel$ bounded at the Curie value of 4.6, and only $A$ is fitted over the whole field range.
The fit runs to the bound, with $T^*_\mathrm{YS} = 2.0$~K, and the resulting curve saturates while the data do not, leaving a standard error of 2.8 to 2.9~pp for windows ending between 4 and 6~T.
Left free, the low-field fit needs $g_\parallel = 7.1$, and 6.5 to 8.2 as the window ends between 4 and 6~T, well above the measured 4.24 to 4.59.
With $A$ and $B_1$ both free over the whole field range (Supplementary Fig.~\ref{sfig:model-comparison}), Yosida still saturates and leaves 3 to 16 times the standard error of $\rho_\mathrm{Kondo,SM} - \rho(B^2)$.
Its $B_1$ stays between 4.7 and 5.9~T from 0.44 to 4.3~K, where $B_1 \propto T + T^*_\mathrm{YS}$ should grow by a factor of 2.6.

\begin{figure}
    \centering
    \includegraphics[width=\figwidth]{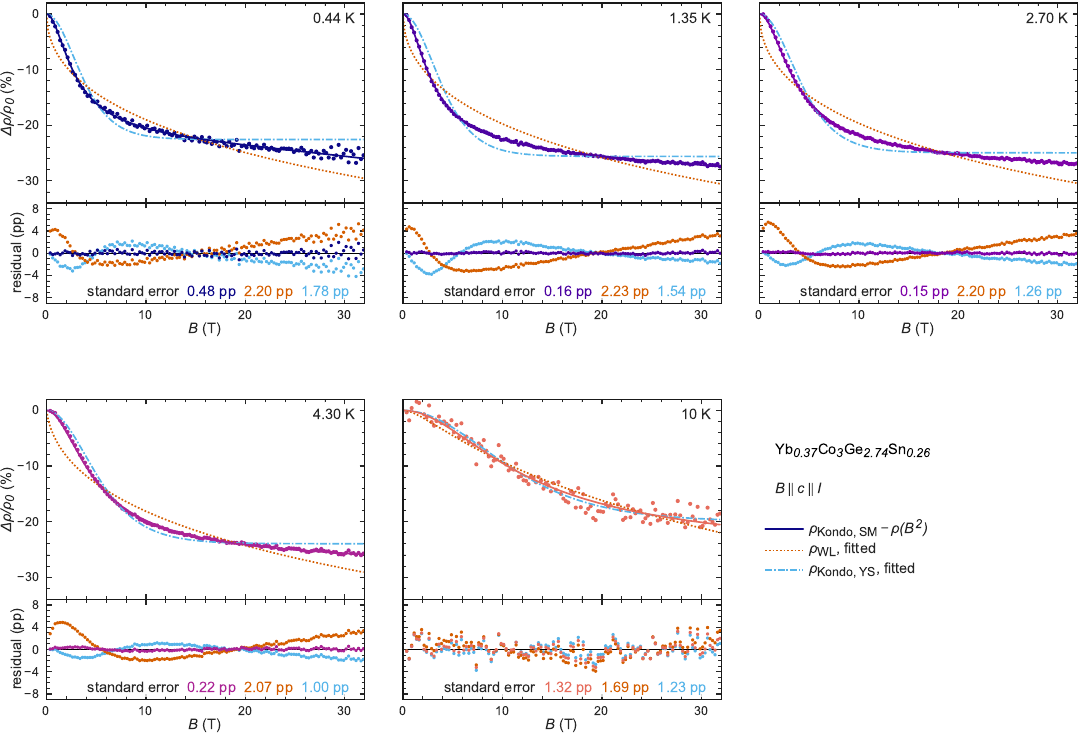}
    \caption{Weak localization and Yosida's spin-disorder term fitted freely to every isotherm of main-text Fig.~3b, as in Supplementary Fig.~\ref{sfig:model-comparison}.
    Each isotherm of \crystal{RM230}{formula} is shown with the fit $\rho_\mathrm{Kondo,SM} - \rho(B^2)$ of main-text Fig.~3b, with $\rho_\mathrm{WL}$ \autocite{Kawabata_1980} with free amplitude and dephasing field, and with $\rho_\mathrm{Kondo,YS}$ \autocite{Yosida_1957} with free amplitude and field scale $B_1$, each fitted from 0.3 to 32~T without $\rho_\mathrm{Kondo,SM}$.
    At 4.3 and 10~K the fit resolves no channel, so $\rho_\mathrm{Kondo,SM} - \rho(B^2)$ is the Schlottmann curve alone.
    Residuals are in percentage points (pp), and the standard errors are as defined in Section~\ref{ssec:standard-error}.}
    \label{sfig:model-comparison-rm230}
\end{figure}

\crystal{RM230}{formula} behaves the same way (Supplementary Fig.~\ref{sfig:model-comparison-rm230}).
With $g_\parallel$ bounded at its Curie value of 4.8, the curve leaves 3.1~pp at 0.44~K, six times the 0.48~pp of $\rho_\mathrm{Kondo,SM} - \rho(B^2)$, and left free the low-field fit needs $g_\parallel = 5.7$ to 6.3.
Fitted freely, Yosida leaves 3.7 to 9.3 times that standard error from 0.44 to 4.3~K, and its $B_1$ rises by a factor of 1.4 where $T + T^*_\mathrm{YS}$ grows by 2.2.

\subsection{Current jetting and contact pairs}
\label{ssec:current-jetting}

\begin{figure}
    \centering
    \includegraphics[width=0.75\textwidth]{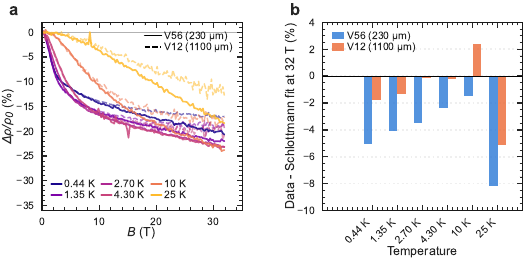}
    \caption{Longitudinal magnetoresistance of \crystal{RM154}{formula} on two contact pairs measured at the same time, for $B \parallel c \parallel I$.
    (a) Isotherms on V56, 230~$\mu$m apart (solid), and on V12, 1100~$\mu$m apart (dashed).
    (b) The data at 32~T less a Schlottmann fit to the field range below 12~T, for each pair and temperature.
    Below 5~K the tail beyond the Schlottmann curve is about three times larger on the narrower pair V56.
    At 10 and 25~K the Kondo field scale approaches the highest measured field, and the model does not separate the two channels well, making the difference somewhat arbitrary.}
    \label{sfig:contact-pairs}
\end{figure}

Current jetting could explain a negative longitudinal magnetoresistance, coming out of the measurement geometry, but it needs high mobility, $\mu B \gg 1$ \autocite{Armitage_etal_2018,Ong_Liang_2021}.
Hall data on nominally stoichiometric \ce{YbCo6Ge6} give an in-plane mobility of 22 to 26~cm$^2$\,V$^{-1}$\,s$^{-1}$ below 20~K \autocite{Ain_etal_2026}, so $\mu B \approx 0.08$ at 32~T for $B \parallel c$, and the majority carriers cannot cause it.
This average does not exclude a small population of high-mobility carriers, but jetting by them should enhance the negative tail on a wider contact pair.
\crystal{RM154}{formula} was measured on two voltage pairs at the same time, V56 and the wider V12 (Methods of the main text).
On V12 the tail is about three times smaller (Supplementary Fig.~\ref{sfig:contact-pairs}): at 0.44~K and 32~T the data lie 1.8~\% below a Schlottmann fit to the field range below 12~T, against 5.1~\% on V56, and the extra channel of Section~\ref{ssec:extra-channel} is resolved only below 3~K.
Current jetting therefore cannot explain the negative LMR.

\subsection{$g$ factors from the different measurement techniques}
\label{ssec:g-factors}

The $g$ factors from the different measurement techniques give different values.
The susceptibility is strongly anisotropic: Curie--Weiss fits between 20 and 250~K (Supplementary Fig.~\ref{sfig:curie-weiss}) give a Weiss temperature of $-3$ to $-4$~K with $B\parallel c$ against $-107$ to $-115$~K $B\perp c$.
For the batch of \crystal{RM154}{formula}, the low-temperature Curie fits give $g_\parallel$ between 4.24 and 4.59 and $g_\perp \approx 1.3$, a ratio of about 3.5, while the field scales of the magnetoresistance give $g_\parallel / g_\perp \approx 2.0$.
The heat capacity of \crystal{RM139}{formula}, which has the same filling, gives $g_\perp = 3.96$ (Section~\ref{ssec:resonant-level}), three times the magnetization value.
A transport ratio of 2.0 together with this $g_\perp$ would need $g_\parallel \approx 8.0$, far above the Curie value.
That the transport ratio matches neither the magnetization nor the heat capacity is a further sign that the $B \parallel c$ field scale of the magnetoresistance is not set by the Yb doublet alone.
We do not yet understand why the heat capacity and the magnetization disagree on $g_\perp$.
The number of doublets does not account for it: with $n$ held at one doublet per Yb the heat capacity still gives $g_\perp = 3.89$, and with $g_\perp$ held at the magnetization value the fit no longer reproduces the Schottky maximum.
At the lower filling of \crystal{RM048}{short} the heat capacity gives $g_\perp = 1.69$, in line with the magnetization of \crystal{RM230}{short}, so the $g$ factor of the doublet may also change with filling.

\subsection{Density-functional calculations}
\label{ssec:dft}

\begin{figure}
    \centering
    \includegraphics[width=\textwidth]{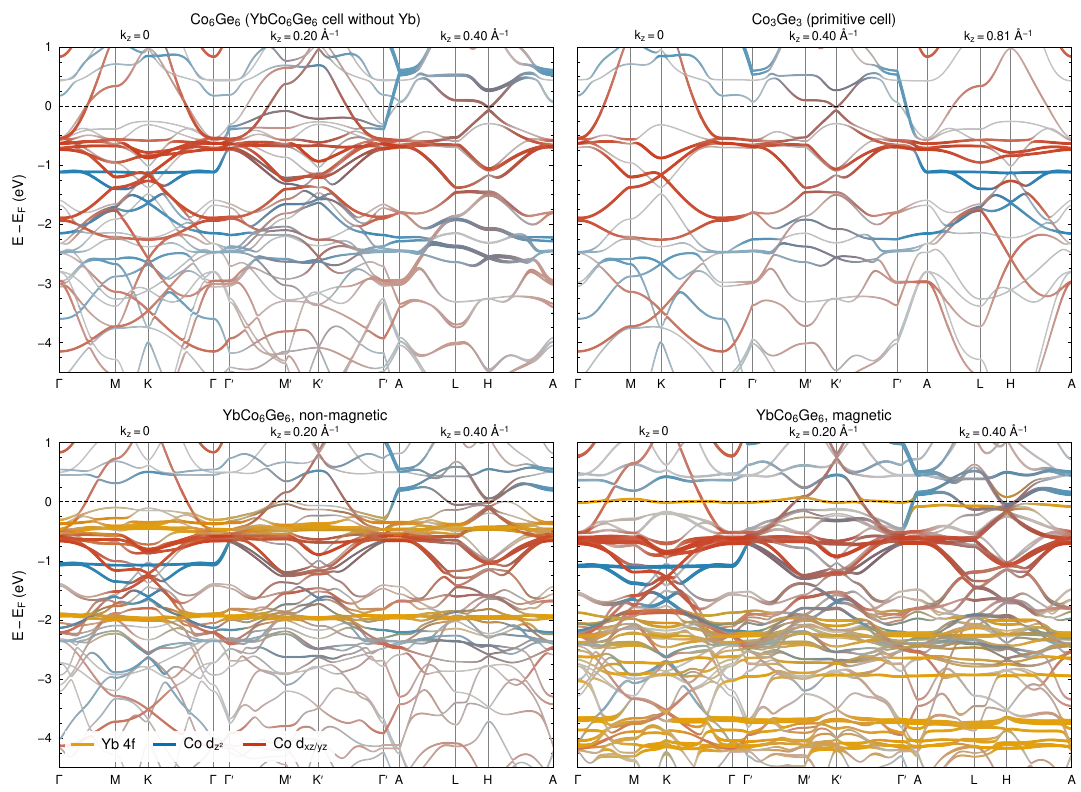}
    \caption{Band structures from density-functional theory on three planes of constant $k_z$: $\Gamma$--M--K--$\Gamma$ at $k_z = 0$, $\Gamma'$--M$'$--K$'$--$\Gamma'$ halfway to the zone boundary, and A--L--H--A on the zone boundary.
    (a) \ce{Co6Ge6}, the \ce{YbCo6Ge6} cell without Yb, which is a doubled \ce{CoSn}-type cell.
    (b) \ce{Co3Ge3}, the primitive \ce{CoSn}-type cell, whose zone boundary lies twice as far out in $k_z$.
    (c) \ce{YbCo6Ge6}, the non-magnetic DFT+$U$+SOC solution.
    (d) \ce{YbCo6Ge6}, the magnetic DFT+$U$+SOC solution, started from $1\,\mu_\mathrm{B}$ on Yb along $c$.
    Panels a and b are PBE with spin-orbit coupling and no $U$.
    Red, blue and orange indicate Co $d_{xz/yz}$, Co $d_{z^2}$ and Yb $4f$ orbital contributions, respectively.}
    \label{sfig:dft}
\end{figure}
To isolate the role of Yb, we studied two CoGe reference systems using plain PBE with spin-orbit coupling (no $U$) and otherwise identical settings:
(i) \ce{Co6Ge6}, obtained from the \ce{YbCo6Ge6} cell by removing Yb and moving the two Ge atoms of the Ge--Ge dimer into the centres of the Co hexagons in the kagome planes, which yields a doubled \ce{CoSn}-type cell with the \ce{YbCo6Ge6} lattice constants (Fig. \ref{sfig:dft}a);
(ii) the primitive \ce{CoSn}-type cell of (i) with $c/2 = 3.902$~\AA{} and three Co and three Ge atoms, for which the Co kagome planes are flattened by 0.013~\AA{} so that the cell can be halved exactly, using an $8\times8\times12$ $k$-point mesh (Fig.~\ref{sfig:dft}b).
All systems converge to non-magnetic ground states; for (i), a ferromagnetic starting configuration with $2\,\mu_\mathrm{B}$ per Co relaxes back to the same non-magnetic state.
Comparing the two cells shows the folding described in the main text.
In the primitive cell (Fig.~\ref{sfig:dft}b) the flat $d_{z^2}$ band near $-1.1$~eV appears only at $k_z = \pi/c = 0.81$~\AA$^{-1}$, along A--L--H, while the $k_z = 0$ plane holds the $d_{xz}/d_{yz}$ bands.
In the doubled cell (Fig.~\ref{sfig:dft}a) the $d_{z^2}$ band folds onto $k_z = 0$ and appears along $\Gamma$--M--K next to the $d_{xz}/d_{yz}$ bands.
Indeed, the lower Dirac-like $d_{xz}/d_{yz}$ band is quadratic around $\Gamma$ with its minimum near $-1.9$~eV in DFT, close to the parabola observed at $\approx -1.7$~eV in ARPES (main-text Fig.~4a,b).
The two flat ARPES features at $\approx -0.5$ and $-0.75$~eV we attribute to the kagome flat bands of $d_{xz}/d_{yz}$ (DFT: $-0.6$~eV) and $d_{z^2}$ (DFT: $-1.1$~eV) character, respectively.
Note that the exact energetic positions of these bands will depend on doping and disorder, which are both not taken realistically into account in our simplified DFT modelling.
Treating the Yb $4f$ states explicitly within DFT+$U$+SOC, the non-magnetic solution yields two spin-orbit-split sets of fully occupied, nearly dispersionless $4f$ bands around $-0.5$ and $-2$~eV (their positions depend on $U$ and $J$).
Two self-consistent solutions of YbCoGe with identical $U$ and $J$ are compared in Fig. \ref{sfig:dft}: a non-magnetic solution obtained from zero initial moments (c), and a magnetic solution obtained from an initial moment of $1\,\mu_\mathrm{B}$ on Yb along the $c$ axis (d).
The total-energy difference between the two solutions changes by less than 1~meV on a $12\times12\times8$ $k$-point mesh.

%% file: My_Library.bib
@misc{_,
  type = {Misc}
}

@article{Ain_etal_2026,
  title = {Disorder {{Driven Anisotropic Nonlinear Hall Effect}} in the {{Kagome Metal YbCo}}{\textsubscript{6}}{{Ge}}{\textsubscript{6}} at {{Room Temperature}}},
  author = {Ain, Qurat Ul and Zhao, Weiyao and Huynh, Thanh Tung and Yin, Yuefeng and Tang, Fang and Fang, Yong and Yan, Junlin and Jasieniak, Jacek and Petersen, Timothy and Liu, Yang and Medhekar, Nikhil V. and Karel, Julie},
  year = 2026,
  month = may,
  doi = {10.1002/advs.77434},
  urldate = {2026-09-08},
  langid = {english}
}

@article{Armitage_etal_2018,
  title = {Weyl and {{Dirac}} Semimetals in Three-Dimensional Solids},
  author = {Armitage, N. P. and Mele, E. J. and Vishwanath, Ashvin},
  year = 2018,
  month = jan,
  journal = {Reviews of Modern Physics},
  volume = {90},
  number = {1},
  pages = {015001},
  publisher = {American Physical Society},
  doi = {10.1103/RevModPhys.90.015001},
  urldate = {2026-08-07}
}

@article{Beal-Monod_Weiner_1968,
  title = {Negative {{Magnetoresistivity}} in {{Dilute Alloys}}},
  author = {{B{\'e}al-Monod}, M.-T. and Weiner, R. A.},
  year = 1968,
  month = jun,
  journal = {Physical Review},
  volume = {170},
  number = {2},
  pages = {552--559},
  publisher = {American Physical Society},
  doi = {10.1103/PhysRev.170.552},
  urldate = {2026-10-15}
}

@article{Bernevig_etal_2022,
  title = {Progress and Prospects in Magnetic Topological Materials},
  author = {Bernevig, B. Andrei and Felser, Claudia and Beidenkopf, Haim},
  year = 2022,
  month = mar,
  journal = {Nature},
  volume = {603},
  number = {7899},
  pages = {41--51},
  publisher = {Nature Publishing Group},
  issn = {1476-4687},
  doi = {10.1038/s41586-021-04105-x},
  urldate = {2026-11-09},
  copyright = {2022 The Author(s), under exclusive licence to Springer Nature Limited},
  langid = {english}
}

@article{Blochl_1994,
  title = {Projector Augmented-Wave Method},
  author = {Bl{\"o}chl, P. E.},
  year = 1994,
  month = dec,
  journal = {Physical Review B},
  volume = {50},
  number = {24},
  pages = {17953--17979},
  publisher = {American Physical Society},
  doi = {10.1103/PhysRevB.50.17953},
  urldate = {2026-11-02}
}

@article{Braun_Hlukhyy_2023,
  title = {Structural Order-Disorder in {{CaFe}}{\textsubscript{6}}{{Ge}}{\textsubscript{6}} and {{Ca}}{\textsubscript{1-x}}{{Co}}{\textsubscript{6}}{{Ge}}{\textsubscript{6}}},
  author = {Braun, Thomas and Hlukhyy, Viktor},
  year = 2023,
  journal = {Journal of Solid State Chemistry},
  volume = {318},
  pages = {123742},
  issn = {0022-4596},
  doi = {10.1016/j.jssc.2022.123742}
}

@article{Bravo_etal_2023,
  title = {Crystal {{Growth}} and {{Physical Properties}} of {{Hybrid CoSn}}--{{YCo}}{\textsubscript{6}}{{Ge}}{\textsubscript{6}} {{Structure Type Ln}}{\textsubscript{{\emph{x}}}}{{Co}}{\textsubscript{3}}({{Ge}}{\textsubscript{1-{\emph{y}}}}{{Sn}}{\textsubscript{{\emph{y}}}}){\textsubscript{3}} ({{Ln}} = {{Y}}, {{Gd}})},
  author = {Bravo, Mois{\'e}s and McCandless, Gregory T. and Baumbach, Ryan E. and Wang, Yaojia and Ali, Mazhar N. and Chan, Julia Y.},
  year = 2023,
  month = oct,
  journal = {Inorganic Chemistry},
  volume = {62},
  number = {44},
  pages = {18049--18055},
  issn = {0020-1669},
  doi = {10.1021/acs.inorgchem.3c02172},
  urldate = {2026-08-07}
}

@article{Canfield_etal_2016,
  title = {Use of Frit-Disc Crucibles for Routine and Exploratory Solution Growth of Single Crystalline Samples},
  author = {Canfield, Paul C. and Kong, Tai and Kaluarachchi, Udhara S. and Jo, Na Hyun},
  year = 2016,
  month = jan,
  journal = {Philosophical Magazine},
  volume = {96},
  number = {1},
  pages = {84--92},
  publisher = {Taylor \& Francis},
  issn = {1478-6435},
  doi = {10.1080/14786435.2015.1122248},
  urldate = {2026-10-05}
}

@article{Canfield_Fisk_1992,
  title = {Growth of Single Crystals from Metallic Fluxes},
  author = {Canfield, P. C. and Fisk, Z.},
  year = 1992,
  month = jun,
  journal = {Philosophical Magazine B},
  volume = {65},
  number = {6},
  pages = {1117--1123},
  publisher = {Taylor \& Francis},
  issn = {1364-2812},
  doi = {10.1080/13642819208215073},
  urldate = {2026-10-05}
}

@article{Cano_etal_2017,
  title = {Chiral Anomaly Factory: {{Creating Weyl}} Fermions with a Magnetic Field},
  shorttitle = {Chiral Anomaly Factory},
  author = {Cano, Jennifer and Bradlyn, Barry and Wang, Zhijun and Hirschberger, Max and Ong, N. P. and Bernevig, B. A.},
  year = 2017,
  month = apr,
  journal = {Physical Review B},
  volume = {95},
  number = {16},
  pages = {161306},
  publisher = {American Physical Society},
  doi = {10.1103/PhysRevB.95.161306},
  urldate = {2026-10-20}
}

@misc{Cheng_etal_2024,
  title = {Pressure-{{Dependent Magnetotransport Measurement}} in {{Kagome Metal Yb}}{\textsubscript{0.5}}{{Co}}{\textsubscript{3}}{{Ge}}{\textsubscript{3}}},
  author = {Cheng, Zhiyuan and Wang, Yaojia and Wu, Heng and Ali, Mazhar N. and Chan, Julia Y. and Bhattacharyya, Semonti},
  year = 2024,
  month = oct,
  number = {arXiv:2410.23872},
  eprint = {2410.23872},
  primaryclass = {cond-mat.str-el},
  publisher = {arXiv},
  doi = {10.48550/arXiv.2410.23872},
  urldate = {2026-08-06},
  archiveprefix = {arXiv}
}

@article{Chorkendorff_etal_1985,
  title = {The {{Yb-Ni}} Interface Studied with Photoemission Spectroscopy},
  author = {Chorkendorff, I. and Onsgaard, J. and {Schmidt-May}, J. and Nyholm, R.},
  year = 1985,
  month = sep,
  journal = {Surface Science},
  volume = {160},
  number = {2},
  pages = {587--598},
  issn = {0039-6028},
  doi = {10.1016/0039-6028(85)90796-4},
  langid = {english}
}

@article{Ding_etal_2021,
  title = {Crossover from {{Kondo}} to {{Fermi-Liquid Behavior Induced}} by {{High Magnetic Field}} in {{1T-VTe}}{\textsubscript{2}} {{Single Crystals}}},
  author = {Ding, Xiaxin and Xing, Jie and Li, Gang and Balicas, Luis and Gofryk, Krzysztof and Wen, Hai-Hu},
  year = 2021,
  month = mar,
  journal = {Physical Review B},
  volume = {103},
  number = {12},
  pages = {125115},
  publisher = {American Physical Society},
  doi = {10.1103/PhysRevB.103.125115},
  urldate = {2026-10-15}
}

@article{Ding_etal_2026,
  title = {Spin Order and Excitations in the Collinear Kagome Antiferromagnet {{Y}}{\textsubscript{0.5}}{{Fe}}{\textsubscript{3}}{{Sn}}{\textsubscript{3}}},
  author = {Ding, Jin and Liu, Yang and Pan, Feihao and Zhu, Yinghao and Su, Yixi and Nikitin, Stanislav E. and Hansen, Ursula B. and Enderle, Mechthild and Lu, Zhilun and Song, Linxuan and Lau, Yong-Chang and Ma, Xiaoyan and Liu, Enke and Luo, Huiqian},
  year = 2026,
  journal = {Physical Review B},
  volume = {114},
  number = {20},
  pages = {204418},
  issn = {2469-9950},
  doi = {10.1103/mjd4-ylmx}
}

@article{Doniach_1977,
  title = {The {{Kondo}} Lattice and Weak Antiferromagnetism},
  author = {Doniach, S.},
  year = 1977,
  month = jul,
  journal = {Physica B+C},
  volume = {91},
  pages = {231--234},
  issn = {0378-4363},
  doi = {10.1016/0378-4363(77)90190-5},
  urldate = {2025-09-04}
}

@article{Dzsaber_etal_2021,
  title = {Giant Spontaneous {{Hall}} Effect in a Nonmagnetic {{Weyl}}--{{Kondo}} Semimetal},
  author = {Dzsaber, Sami and Yan, Xinlin and Taupin, Mathieu and Eguchi, Gaku and Prokofiev, Andrey and Shiroka, Toni and Blaha, Peter and Rubel, Oleg and Grefe, Sarah E. and Lai, Hsin-Hua and Si, Qimiao and Paschen, Silke},
  year = 2021,
  month = feb,
  journal = {Proceedings of the National Academy of Sciences},
  volume = {118},
  number = {8},
  pages = {e2013386118},
  publisher = {Proceedings of the National Academy of Sciences},
  doi = {10.1073/pnas.2013386118},
  urldate = {2026-09-17}
}

@article{Fau_etal_2026,
  title = {Effect of {{Co}} Doping on Structural and Magnetic Order in Kagome Magnet {{TbFe}}{\textsubscript{6-{\emph{x}}}}{{Co}}{\textsubscript{{\emph{x}}}}{{Ge}}{\textsubscript{6}} (Nominal {\emph{x}} = 0-6)},
  author = {Fau, Noah J. and Meduri, Rahul and Das, Atreyee and Baumbach, Ryan E. and McCandless, Gregory T. and Chan, Julia Y.},
  year = 2026,
  month = sep,
  journal = {Physical Review Materials},
  volume = {10},
  number = {9},
  pages = {095002},
  publisher = {American Physical Society},
  doi = {10.1103/my7d-txwd},
  urldate = {2026-09-21}
}

@article{Goncalves_etal_1994,
  title = {{{UFe}}{\textsubscript{6}}{{Ge}}{\textsubscript{6}}: A New Ternary Magnetic Compound},
  author = {Gon{\c c}alves, A.P. and Waerenborgh, J.C. and Bonfait, G. and Amaro, A. and Godinho, M.M. and Almeida, M. and Spirlet, J.C.},
  year = 1994,
  journal = {Journal of Alloys and Compounds},
  volume = {204},
  number = {1-2},
  pages = {59--64},
  issn = {0925-8388},
  doi = {10.1016/0925-8388(94)90071-X}
}

@article{Grefe_etal_2020,
  title = {Weyl-{{Kondo}} Semimetals in Nonsymmorphic Systems},
  author = {Grefe, Sarah E. and Lai, Hsin-Hua and Paschen, Silke and Si, Qimiao},
  year = 2020,
  month = feb,
  journal = {Physical Review B},
  volume = {101},
  number = {7},
  pages = {075138},
  publisher = {American Physical Society},
  doi = {10.1103/PhysRevB.101.075138},
  urldate = {2026-06-01}
}

@incollection{Grewe_Steglich_1991,
  title = {Chapter 97 {{Heavy}} Fermions},
  booktitle = {Handbook on the {{Physics}} and {{Chemistry}} of {{Rare Earths}}},
  author = {Grewe, Norbert and Steglich, Frank},
  year = 1991,
  month = jan,
  volume = {14},
  pages = {343--474},
  publisher = {Elsevier},
  doi = {10.1016/S0168-1273(05)80103-5},
  urldate = {2025-09-11}
}

@article{Guo_etal_2018,
  title = {Evidence for {{Weyl}} Fermions in a Canonical Heavy-Fermion Semimetal {{YbPtBi}}},
  author = {Guo, C. Y. and Wu, F. and Wu, Z. Z. and Smidman, M. and Cao, C. and Bostwick, A. and Jozwiak, C. and Rotenberg, E. and Liu, Y. and Steglich, F. and Yuan, H. Q.},
  year = 2018,
  month = nov,
  journal = {Nature Communications},
  volume = {9},
  number = {1},
  pages = {4622},
  publisher = {Nature Publishing Group},
  issn = {2041-1723},
  doi = {10.1038/s41467-018-06782-1},
  urldate = {2026-07-28},
  copyright = {2018 The Author(s)},
  langid = {english}
}

@article{Guo_etal_2023a,
  title = {Triangular {{Kondo Lattice}} in {{YbV}}{\textsubscript{6}}{{Sn}}{\textsubscript{6}} and {{Its Quantum Critical Behavior}} in a {{Magnetic Field}}},
  author = {Guo, Kaizhen and Ye, Junyao and Guan, Shuyue and Jia, Shuang},
  year = 2023,
  month = may,
  journal = {Physical Review B},
  volume = {107},
  number = {20},
  pages = {205151},
  publisher = {American Physical Society},
  doi = {10.1103/PhysRevB.107.205151},
  urldate = {2026-02-25}
}

@article{Hirschberger_etal_2016,
  title = {The Chiral Anomaly and Thermopower of {{Weyl}} Fermions in the Half-{{Heusler GdPtBi}}},
  author = {Hirschberger, Max and Kushwaha, Satya and Wang, Zhijun and Gibson, Quinn and Liang, Sihang and Belvin, Carina A. and Bernevig, B. A. and Cava, R. J. and Ong, N. P.},
  year = 2016,
  month = nov,
  journal = {Nature Materials},
  volume = {15},
  number = {11},
  pages = {1161--1165},
  publisher = {Nature Publishing Group},
  issn = {1476-4660},
  doi = {10.1038/nmat4684},
  urldate = {2026-10-20},
  copyright = {2016 Springer Nature Limited},
  langid = {english}
}

@article{HjorthLarsen_etal_2017,
  title = {The Atomic Simulation Environment---a {{Python}} Library for Working with Atoms},
  author = {Hjorth Larsen, Ask and J{\o}rgen Mortensen, Jens and Blomqvist, Jakob and Castelli, Ivano E and Christensen, Rune and Du{\l}ak, Marcin and Friis, Jesper and Groves, Michael N and Hammer, Bj{\o}rk and Hargus, Cory and Hermes, Eric D and Jennings, Paul C and Bjerre Jensen, Peter and Kermode, James and Kitchin, John R and Leonhard Kolsbjerg, Esben and Kubal, Joseph and Kaasbjerg, Kristen and Lysgaard, Steen and Bergmann Maronsson, J{\'o}n and Maxson, Tristan and Olsen, Thomas and Pastewka, Lars and Peterson, Andrew and Rostgaard, Carsten and Schi{\o}tz, Jakob and Sch{\"u}tt, Ole and Strange, Mikkel and Thygesen, Kristian S and Vegge, Tejs and Vilhelmsen, Lasse and Walter, Michael and Zeng, Zhenhua and Jacobsen, Karsten W},
  year = 2017,
  month = jun,
  journal = {Journal of Physics: Condensed Matter},
  volume = {29},
  number = {27},
  pages = {273002},
  publisher = {IOP Publishing},
  issn = {0953-8984},
  doi = {10.1088/1361-648X/aa680e},
  urldate = {2026-11-02},
  langid = {english}
}

@article{Hobbs_etal_2000,
  title = {Fully Unconstrained Noncollinear Magnetism within the Projector Augmented-Wave Method},
  author = {Hobbs, D. and Kresse, G. and Hafner, J.},
  year = 2000,
  month = nov,
  journal = {Physical Review B},
  volume = {62},
  number = {17},
  pages = {11556--11570},
  publisher = {American Physical Society},
  doi = {10.1103/PhysRevB.62.11556},
  urldate = {2026-11-02}
}

@article{johnston_hoffmann_1990,
  title = {The Kagom\'e Net: {{Band}} Theoretical and Topological Aspects},
  shorttitle = {The Kagom\'e Net},
  author = {Johnston, Roy L. and Hoffmann, Roald},
  year = 1990,
  month = jan,
  journal = {Polyhedron},
  volume = {9},
  number = {15-16},
  pages = {1901--1911},
  issn = {02775387},
  doi = {10.1016/S0277-5387(00)84002-4},
  urldate = {2022-09-23},
  langid = {english}
}

@article{Kawabata_1980,
  title = {Theory of Negative Magnetoresistance in Three-Dimensional Systems},
  author = {Kawabata, A.},
  year = 1980,
  month = may,
  journal = {Solid State Communications},
  volume = {34},
  number = {6},
  pages = {431--432},
  issn = {0038-1098},
  doi = {10.1016/0038-1098(80)90644-4},
  urldate = {2026-09-22}
}

@article{Kohler_etal_2008,
  title = {Energy {{Scales}} of {{Lu}}{\textsubscript{1-x}}{{Yb}}{\textsubscript{x}}{{Rh}}{\textsubscript{2}}{{Si}}{\textsubscript{2}} by {{Means}} of {{Thermopower Investigations}}},
  author = {K{\"o}hler, U. and Oeschler, N. and Steglich, F. and Maquilon, S. and Fisk, Z.},
  year = 2008,
  month = mar,
  journal = {Physical Review B},
  volume = {77},
  number = {10},
  pages = {104412},
  publisher = {American Physical Society},
  doi = {10.1103/PhysRevB.77.104412},
  urldate = {2026-08-07}
}

@article{Kondo_1964,
  title = {Resistance Minimum in Dilute Magnetic Alloys},
  author = {Kondo, Jun},
  year = 1964,
  journal = {Progress of Theoretical Physics},
  volume = {32},
  number = {1},
  pages = {37--49},
  doi = {10.1143/PTP.32.37}
}

@article{Kresse_Furthmuller_1996,
  title = {Efficient Iterative Schemes for Ab Initio Total-Energy Calculations Using a Plane-Wave Basis Set},
  author = {Kresse, G. and Furthm{\"u}ller, J.},
  year = 1996,
  month = oct,
  journal = {Physical Review B},
  volume = {54},
  number = {16},
  pages = {11169--11186},
  publisher = {American Physical Society},
  doi = {10.1103/PhysRevB.54.11169},
  urldate = {2026-11-02}
}

@article{Kresse_Furthmuller_1996a,
  title = {Efficiency of Ab-Initio Total Energy Calculations for Metals and Semiconductors Using a Plane-Wave Basis Set},
  author = {Kresse, G. and Furthm{\"u}ller, J.},
  year = 1996,
  month = jul,
  journal = {Computational Materials Science},
  volume = {6},
  number = {1},
  pages = {15--50},
  issn = {0927-0256},
  doi = {10.1016/0927-0256(96)00008-0},
  urldate = {2026-11-02}
}

@article{Kresse_Joubert_1999,
  title = {From Ultrasoft Pseudopotentials to the Projector Augmented-Wave Method},
  author = {Kresse, G. and Joubert, D.},
  year = 1999,
  month = jan,
  journal = {Physical Review B},
  volume = {59},
  number = {3},
  pages = {1758--1775},
  publisher = {American Physical Society},
  doi = {10.1103/PhysRevB.59.1758},
  urldate = {2026-11-02}
}

@article{Lai_etal_2018,
  title = {Weyl--{{Kondo}} Semimetal in Heavy-Fermion Systems},
  author = {Lai, Hsin-Hua and Grefe, Sarah E. and Paschen, Silke and Si, Qimiao},
  year = 2018,
  month = jan,
  journal = {Proceedings of the National Academy of Sciences},
  volume = {115},
  number = {1},
  pages = {93--97},
  publisher = {Proceedings of the National Academy of Sciences},
  doi = {10.1073/pnas.1715851115},
  urldate = {2026-10-05}
}

@article{Lee_etal_2026,
  title = {Coexisting {{Kagome}} and {{Heavy Fermion Flat Bands}} in {{YbCr}}{\textsubscript{6}}{{Ge}}{\textsubscript{6}}},
  author = {Lee, Hanoh and Lyi, Churlhi and Lee, Taehee and Na, Hyeonhui and Kim, Jinyoung and Lee, Sangjae and Kim, Younsik and Rajapitamahuni, Anil and Kundu, Asish K. and Vescovo, Elio and Park, Byeong-Gyu and Kim, Changyoung and Ahn, Charles H. and Walker, Frederick J. and Oh, Ji Seop and Jang, Bo Gyu and Kim, Youngkuk and Sohn, Byungmin and Park, Tuson},
  year = 2026,
  month = mar,
  journal = {Nature Communications},
  publisher = {Nature Publishing Group},
  issn = {2041-1723},
  doi = {10.1038/s41467-026-70958-3},
  urldate = {2026-03-31},
  copyright = {2026 The Author(s)},
  langid = {english}
}

@article{Li_etal_2025a,
  title = {Generating {{Ferromagnetic Kagom\'e Metal}} by {{Magnetic Phase Boundary Mapping}} in the {{YFe}}{\textsubscript{6}}{{Ge}}{\textsubscript{6}}--{{YCo}}{\textsubscript{6}}{{Ge}}{\textsubscript{6}} {{System}}},
  author = {Li, Victoria M. and Adams, Milo and Shatruk, Michael},
  year = 2025,
  journal = {Chemistry of Materials},
  volume = {37},
  number = {1},
  pages = {247--254},
  issn = {0897-4756},
  doi = {10.1021/acs.chemmater.4c02382}
}

@article{Liechtenstein_etal_1995,
  title = {Density-Functional Theory and Strong Interactions: {{Orbital}} Ordering in {{Mott-Hubbard}} Insulators},
  shorttitle = {Density-Functional Theory and Strong Interactions},
  author = {Liechtenstein, A. I. and Anisimov, V. I. and Zaanen, J.},
  year = 1995,
  month = aug,
  journal = {Physical Review B},
  volume = {52},
  number = {8},
  pages = {R5467-R5470},
  publisher = {American Physical Society},
  doi = {10.1103/PhysRevB.52.R5467},
  urldate = {2023-03-06}
}

@article{Liu_etal_2023b,
  title = {Structural {{Determination}}, {{Unstable Antiferromagnetism}} and {{Transport Properties}} of {{Fe-Kagome Y}}{\textsubscript{0.5}}{{Fe}}{\textsubscript{3}}{{Sn}}{\textsubscript{3}} {{Single Crystals}}},
  author = {Liu, Yang and Lyu, Meng and Liu, Junyan and Zhang, Shen and Yang, Jinying and Du, Zhiwei and Wang, Binbin and Wei, Hongxiang and Liu, Enke},
  year = 2023,
  journal = {Chinese Physics Letters},
  volume = {40},
  number = {4},
  pages = {047102},
  issn = {0256-307X},
  doi = {10.1088/0256-307X/40/4/047102}
}

@article{Lou_etal_2025b,
  title = {Strongly {{Entangled Kondo}} and {{Kagome Lattices}} and the {{Emergent Magnetic Ground State}} in {{Heavy-Fermion Kagome Metal YbV}}{\textsubscript{6}}{{Sn}}{\textsubscript{6}}},
  author = {Lou, Rui and Mende, Max and Vocaturo, Riccardo and Zhang, Hao and Dong, Qingxin and Li, Man and Ding, Pengfei and Cheng, Erjian and Liao, Zhiguang and Zhang, Yu and Lin, Junfa and Firouzmandi, Reza and Kocsis, Vilmos and Corredor, Laura T. and Prots, Yurii and Suvorov, Oleksandr and Jana, Anupam and Fujii, Jun and Vobornik, Ivana and Janson, Oleg and Zhu, Wenliang and {van den Brink}, Jeroen and Krellner, Cornelius and Pan, Minghu and Wang, Bosen and Xia, Tianlong and Cheng, Jinguang and Wang, Shancai and Felser, Claudia and B{\"u}chner, Bernd and Borisenko, Sergey and Yu, Rong and Vyalikh, Denis V. and Fedorov, Alexander},
  year = 2025,
  month = sep,
  journal = {Physical Review Letters},
  volume = {135},
  number = {14},
  pages = {146902},
  publisher = {American Physical Society},
  doi = {10.1103/sqbn-vtcj},
  urldate = {2026-02-25}
}

@article{Meduri_etal_2025,
  title = {Evolution of {{Structural Order}} and {{Magnetic Anisotropy}} in {{Yb}}{\textsubscript{0.5}}({{Co}}{\textsubscript{1-{\emph{x}}}}{{Fe}}{\textsubscript{{\emph{x}}}}){\textsubscript{3}}{{Ge}}{\textsubscript{3}} through {{Doping}} of a {{Kagome Lattice}}},
  author = {Meduri, Rahul and Plata, Mario A. and McCandless, Gregory T. and Schundelmier, Benny C. and Ghafoor, Mehak and Wei, Kaya and Chan, Julia Y.},
  year = 2025,
  month = mar,
  journal = {Chemistry of Materials},
  volume = {37},
  number = {6},
  pages = {2302--2313},
  issn = {0897-4756},
  doi = {10.1021/acs.chemmater.5c00119},
  urldate = {2026-08-07}
}

@article{Nielsen_Ninomiya_1983,
  title = {The {{Adler-Bell-Jackiw}} Anomaly and {{Weyl}} Fermions in a Crystal},
  author = {Nielsen, H. B. and Ninomiya, Masao},
  year = 1983,
  month = nov,
  journal = {Physics Letters B},
  volume = {130},
  number = {6},
  pages = {389--396},
  issn = {0370-2693},
  doi = {10.1016/0370-2693(83)91529-0},
  urldate = {2026-08-07}
}

@article{Ohno_2008,
  title = {{{XPS}} Studies of the Intermediate Valence State of {{Yb}} in ({{YbS}}){\textsubscript{1.25}}{{CrS}}{\textsubscript{2}}},
  author = {Ohno, Youichi},
  year = 2008,
  month = sep,
  journal = {Journal of Electron Spectroscopy and Related Phenomena},
  volume = {165},
  number = {1-3},
  pages = {1--4},
  issn = {0368-2048},
  doi = {10.1016/j.elspec.2008.05.009},
  langid = {english}
}

@article{Ong_Liang_2021,
  title = {Experimental Signatures of the Chiral Anomaly in {{Dirac}}--{{Weyl}} Semimetals},
  author = {Ong, N. P. and Liang, Sihang},
  year = 2021,
  month = jun,
  journal = {Nature Reviews Physics},
  volume = {3},
  number = {6},
  pages = {394--404},
  publisher = {Nature Publishing Group},
  issn = {2522-5820},
  doi = {10.1038/s42254-021-00310-9},
  urldate = {2026-07-29},
  copyright = {2021 Springer Nature Limited},
  langid = {english}
}

@article{Onuki_etal_1985c,
  title = {Kondo {{Lattice Formation}} in {{Ce}}{\textsubscript{{\emph{x}}}}{{La}}{\textsubscript{1-{\emph{x}}}}{{Cu}}{\textsubscript{6}}},
  author = {{\=O}nuki, Yoshichika and Shimizu, Yoshiaki and Nishihara, Munekazu and Machii, Yohichi and Komatsubara, Takemi},
  year = 1985,
  month = may,
  journal = {Journal of the Physical Society of Japan},
  volume = {54},
  number = {5},
  pages = {1964--1974},
  publisher = {The Physical Society of Japan},
  issn = {0031-9015},
  doi = {10.1143/JPSJ.54.1964},
  urldate = {2026-08-07}
}

@article{Perdew_etal_1996,
  title = {Generalized {{Gradient Approximation Made Simple}}},
  author = {Perdew, John P. and Burke, Kieron and Ernzerhof, Matthias},
  year = 1996,
  month = oct,
  journal = {Physical Review Letters},
  volume = {77},
  number = {18},
  pages = {3865--3868},
  publisher = {American Physical Society},
  doi = {10.1103/PhysRevLett.77.3865},
  urldate = {2026-11-02}
}

@article{Petrovic_etal_2023,
  title = {The {{Kondo Effect}} in {{Ce}}{\textsubscript{{\emph{x}}}}{{LaLuScY}} ({\emph{x}} = 0.05--1.0) {{High-Entropy Alloys}}},
  author = {Petrovi{\'c}, Julia and Vrtnik, Stanislav and Jelen, Andreja and Ko{\v z}elj, Primo{\v z} and Luzar, Jo{\v z}e and Mihor, Peter and Hu, Qiang and Wencka, Magdalena and Ambro{\v z}i{\v c}, Bojan and Meden, Anton and Dra{\v z}i{\'c}, Goran and Guo, Sheng and Dolin{\v s}ek, Janez},
  year = 2023,
  month = jan,
  journal = {Materials},
  volume = {16},
  number = {24},
  pages = {7575},
  publisher = {Multidisciplinary Digital Publishing Institute},
  issn = {1996-1944},
  doi = {10.3390/ma16247575},
  urldate = {2026-08-07},
  copyright = {http://creativecommons.org/licenses/by/3.0/},
  langid = {english}
}

@article{Schlottmann_1983,
  title = {Bethe-{{Ansatz Solution}} of the {{Ground-State}} of the {{SU}}(2{\emph{j}}+1) {{Kondo}} ({{Coqblin-Schrieffer}}) {{Model}}: {{Magnetization}}, {{Magnetoresistance}} and {{Universality}}},
  shorttitle = {Bethe-{{Ansatz}} Solution of the Ground-State of {{theSU}} (2j+1) {{Kondo}} ({{Coqblin-Schrieffer}}) Model},
  author = {Schlottmann, P.},
  year = 1983,
  month = sep,
  journal = {Zeitschrift f\"ur Physik B Condensed Matter},
  volume = {51},
  number = {3},
  pages = {223--235},
  issn = {1431-584X},
  doi = {10.1007/BF01307678},
  urldate = {2025-06-19},
  langid = {english}
}

@article{Schmidt_etal_2005,
  title = {X-Ray Photoemission of {{YbInCu}}{\textsubscript{4}}},
  author = {Schmidt, S. and H{\"u}fner, S. and Reinert, F. and Assmus, W.},
  year = 2005,
  month = may,
  journal = {Physical Review B},
  volume = {71},
  number = {19},
  pages = {195110},
  issn = {1098-0121, 1550-235X},
  doi = {10.1103/physrevb.71.195110},
  langid = {english}
}

@article{Schotte_Schotte_1975,
  title = {Interpretation of {{Kondo}} Experiments in a Magnetic Field},
  author = {Schotte, K. D. and Schotte, U.},
  year = 1975,
  month = nov,
  journal = {Physics Letters A},
  volume = {55},
  number = {1},
  pages = {38--40},
  issn = {0375-9601},
  doi = {10.1016/0375-9601(75)90386-2},
  urldate = {2026-02-09}
}

@article{Slebarski_etal_2013,
  title = {Evolution from a {{Magnetically Correlated State}} to a {{Single Impurity State}} in {{Heavy Fermion System Ce}}{\textsubscript{3-{\emph{x}}}}{{La}}{\textsubscript{{\emph{x}}}}{{Co}}{\textsubscript{4}}{{Sn}}{\textsubscript{13}}},
  author = {{\'S}lebarski, A. and Fija{\l}kowski, M. and Goraus, J.},
  year = 2013,
  journal = {physica status solidi (b)},
  volume = {250},
  number = {3},
  pages = {472--475},
  issn = {1521-3951},
  doi = {10.1002/pssb.201200571},
  urldate = {2026-08-07},
  copyright = {Copyright \copyright{} 2013 WILEY-VCH Verlag GmbH \& Co. KGaA, Weinheim},
  langid = {english}
}

@article{Soluyanov_etal_2015,
  title = {Type-{{II Weyl}} Semimetals},
  author = {Soluyanov, Alexey A. and Gresch, Dominik and Wang, Zhijun and Wu, QuanSheng and Troyer, Matthias and Dai, Xi and Bernevig, B. Andrei},
  year = 2015,
  month = nov,
  journal = {Nature},
  volume = {527},
  number = {7579},
  pages = {495--498},
  issn = {0028-0836},
  doi = {10.1038/nature15768}
}

@article{Son_Spivak_2013,
  title = {Chiral Anomaly and Classical Negative Magnetoresistance of {{Weyl}} Metals},
  author = {Son, D. T. and Spivak, B. Z.},
  year = 2013,
  month = sep,
  journal = {Physical Review B},
  volume = {88},
  number = {10},
  pages = {104412},
  publisher = {American Physical Society},
  doi = {10.1103/PhysRevB.88.104412},
  urldate = {2026-08-06}
}

@article{Steiner_etal_2016,
  title = {Calculation of the Magnetic Anisotropy with Projected-Augmented-Wave Methodology and the Case Study of Disordered {{Fe}}{\textsubscript{1-{\emph{x}}}}{{Co}}{\textsubscript{{\emph{x}}}} Alloys},
  author = {Steiner, Soner and Khmelevskyi, Sergii and Marsmann, Martijn and Kresse, Georg},
  year = 2016,
  month = jun,
  journal = {Physical Review B},
  volume = {93},
  number = {22},
  pages = {224425},
  publisher = {American Physical Society},
  doi = {10.1103/PhysRevB.93.224425},
  urldate = {2026-11-02}
}

@article{Stewart_1984a,
  title = {Heavy-Fermion Systems},
  author = {Stewart, G.},
  year = 1984,
  month = oct,
  doi = {10.1103/REVMODPHYS.56.755}
}

@article{Sumiyama_etal_1986,
  title = {Coherent {{Kondo State}} in a {{Dense Kondo Substance}}: {{Ce}}{\textsubscript{{\emph{x}}}}{{La}}{\textsubscript{1-{\emph{x}}}}{{Cu}}{\textsubscript{6}}},
  shorttitle = {Coherent {{Kondo State}} in a {{Dense Kondo Substance}}},
  author = {Sumiyama, Akihiko and Oda, Yasukage and Nagano, Hiroshi and {\=O}nuki, Yoshichika and Shibutani, Kazuyuki and Komatsubara, Takemi},
  year = 1986,
  month = apr,
  journal = {Journal of the Physical Society of Japan},
  volume = {55},
  number = {4},
  pages = {1294--1304},
  publisher = {The Physical Society of Japan},
  issn = {0031-9015},
  doi = {10.1143/JPSJ.55.1294},
  urldate = {2026-08-07}
}

@article{Sun_Steglich_2013,
  title = {Nernst {{Effect}}: {{Evidence}} of {{Local Kondo Scattering}} in {{Heavy Fermions}}},
  shorttitle = {Nernst {{Effect}}},
  author = {Sun, Peijie and Steglich, Frank},
  year = 2013,
  month = may,
  journal = {Physical Review Letters},
  volume = {110},
  number = {21},
  pages = {216408},
  publisher = {American Physical Society},
  doi = {10.1103/PhysRevLett.110.216408},
  urldate = {2026-11-09}
}

@article{Tan_etal_2024,
  title = {Three-Dimensional Higher-Order Saddle-Point-Induced Flatbands in {{Co}}-Based Kagome Metals},
  author = {Tan, Hengxin and Jiang, Yiyang and McCandless, Gregory T. and Chan, Julia Y. and Yan, Binghai},
  year = 2024,
  month = nov,
  journal = {Physical Review Research},
  volume = {6},
  number = {4},
  pages = {043132},
  publisher = {American Physical Society},
  doi = {10.1103/PhysRevResearch.6.043132},
  urldate = {2026-08-06}
}

@article{Tanaka_Kontani_2010,
  title = {Intrinsic Spin and Orbital {{Hall}} Effects in Heavy-Fermion Systems},
  author = {Tanaka, T. and Kontani, H.},
  year = 2010,
  month = jun,
  journal = {Physical Review B},
  volume = {81},
  number = {22},
  pages = {224401},
  publisher = {American Physical Society},
  doi = {10.1103/PhysRevB.81.224401},
  urldate = {2026-11-09}
}

@article{Venturini_etal_2008,
  title = {Structures and {{Crystal Chemistry}} of {{MT}}{\textsubscript{6}}{{X}}{\textsubscript{6}} {{Phases}}, {{Filled Derivatives}} of the {{CoSn-B35 Structure}}},
  author = {{Laboratoire de Chimie du Solide Min\'eral, Universit\'e Henri Poincar\'e-Nancy I, associ\'e au CNRS (UMR 7555)} and Venturini, G. and {Ihou-Mouko}, H. and {Laboratoire de Chimie du Solide Min\'eral, Universit\'e Henri Poincar\'e-Nancy I, associ\'e au CNRS (UMR 7555)} and Lef{\`e}vre, C. and {Max-Planck Institut f\"ur Festk\"orperforschung} and Lidin, S. and {Arrhenius Laboratory, Department of Physical, Inorganic and Structural Chemistry, Stockholm University} and Malaman, B. and {Laboratoire de Chimie du Solide Min\'eral, Universit\'e Henri Poincar\'e-Nancy I, associ\'e au CNRS (UMR 7555)} and Mazet, T. and {Laboratoire de Chimie du Solide Min\'eral, Universit\'e Henri Poincar\'e-Nancy I, associ\'e au CNRS (UMR 7555)} and Tobola, J. and {Faculty of Physics and Nuclear Techniques, AGH University of Science and Technology} and Verni{\`e}re, A. and {Laboratoire de Chimie du Solide Min\'eral, Universit\'e Henri Poincar\'e-Nancy I, associ\'e au CNRS (UMR 7555)}},
  year = 2008,
  journal = {Chemistry of Metals and Alloys},
  volume = {1},
  number = {1},
  pages = {24--33},
  issn = {19988079, 19988087},
  doi = {10.30970/cma1.0009},
  urldate = {2026-08-07},
  langid = {english}
}

@article{Wang_etal_2022,
  title = {Electronic {{Properties}} and {{Phase Transition}} in the {{Kagome Metal Yb}}{\textsubscript{0.5}}{{Co}}{\textsubscript{3}}{{Ge}}{\textsubscript{3}}},
  author = {Wang, Yaojia and McCandless, Gregory T. and Wang, Xiaoping and Thanabalasingam, Kulatheepan and Wu, Heng and Bouwmeester, Damian and {van der Zant}, Herre S. J. and Ali, Mazhar N. and Chan, Julia Y.},
  year = 2022,
  month = aug,
  journal = {Chemistry of Materials},
  volume = {34},
  number = {16},
  pages = {7337--7343},
  issn = {0897-4756},
  doi = {10.1021/acs.chemmater.2c01309},
  urldate = {2026-08-07}
}

@article{Weiland_etal_2020,
  title = {Refine {{Intervention}}: {{Characterizing Disordered Yb}}{\textsubscript{0.5}}{{Co}}{\textsubscript{3}}{{Ge}}{\textsubscript{3}}},
  shorttitle = {Refine {{Intervention}}},
  author = {Weiland, Ashley and Eddy, Lucas J. and McCandless, Gregory T. and Hodovanets, Halyna and Paglione, Johnpierre and Chan, Julia Y},
  year = 2020,
  month = aug,
  journal = {Crystal Growth \& Design},
  volume = {20},
  number = {10},
  pages = {6715--6721},
  issn = {1528-7483},
  doi = {10.1021/acs.cgd.0c00865},
  urldate = {2026-08-07}
}

@article{Winkler_etal_2010,
  title = {Lu and {{La Substituted Ce}}{\textsubscript{3}}{{Pd}}{\textsubscript{20}}{{Si}}{\textsubscript{6}}},
  author = {Winkler, H. and Laumann, S. and Custers, J. and Prokofiev, A. and Paschen, S.},
  year = 2010,
  journal = {physica status solidi (b)},
  volume = {247},
  number = {3},
  pages = {516--519},
  issn = {1521-3951},
  doi = {10.1002/pssb.200983056},
  urldate = {2026-08-07},
  copyright = {Copyright \copyright{} 2010 WILEY-VCH Verlag GmbH \& Co. KGaA, Weinheim},
  langid = {english}
}

@article{Wirth_Steglich_2016,
  title = {Exploring Heavy Fermions from Macroscopic to Microscopic Length Scales},
  author = {Wirth, Steffen and Steglich, Frank},
  year = 2016,
  month = aug,
  journal = {Nature Reviews Materials},
  volume = {1},
  number = {10},
  pages = {16051},
  issn = {2058-8437},
  doi = {10.1038/natrevmats.2016.51},
  urldate = {2026-01-14},
  langid = {english}
}

@article{Yosida_1957,
  title = {Anomalous {{Electrical Resistivity}} and {{Magnetoresistance Due}} to an {{{\emph{s}}}}-{\emph{d}} {{Interaction}} in {{Cu-Mn Alloys}}},
  author = {Yosida, Kei},
  year = 1957,
  month = jul,
  journal = {Physical Review},
  volume = {107},
  number = {2},
  pages = {396--403},
  publisher = {American Physical Society},
  doi = {10.1103/PhysRev.107.396},
  urldate = {2026-09-22}
}
